# Effects of Price Regulations on Service Utilization and Public Insurance Costs: Evidence from Telehealth Parity Laws

*By* Piyush Akimitsu[*]

This study explores Telehealth Parity Laws (TPLs) and their heterogeneous treatment effects by policy type on outpatient utilization and Medicare costs, considering broadband and licensure infrastructure. State-specific legislative framings create varied Price (physician reimbursement) and Cost (consumer expense) control combinations within a quasi-experimental design. Partial equilibrium causal estimates reveal negligible impacts on Medicare enrollment, indicating that Medicare cost shifts stem purely from outpatient utilization changes. Broadband access correlates with increased preventable hospital stays but lower Medicare costs and enrollment. Additionally, the Interstate Licensure Compact increases enrollment among the aged and disabled, possibly addressing previously unmet demand for healthcare services.

---

\* Akimitsu: SUNY at Albany, pgade@albany.edu.

# I  INTRODUCTION AND BACKGROUND

Healthcare in the United States is among the most regulated sectors, with policies governing all facets of service delivery, including telehealth. Since 1995, 40 states and the District of Columbia have enacted Telehealth Parity Laws (TPLs), mandating insurance coverage for telehealth services.[1] These laws gained critical significance during the COVID-19 pandemic, as social distancing measures accelerated telehealth adoption.[2] This shift was supported by the increasing use of the internet for health-related purposes, highlighting telehealth's role in reducing disease exposure, improving access for patients in underserved regions, and delivering cost-effective care comparable to in-person treatment.[3]

The "Coverage Parity" aspect of these laws dictate that if an insurance plan covers a service when delivered face-to-face, it must also cover the same service if provided via telehealth. Additionally, many states that introduced Coverage Parity also enacted "Payment Parity" mandates. These mandates guarantee that telehealth services receive the same reimbursement rates for physicians and entail the same deductible, copay, and insurance costs for patients as in-person services. However, the Payment Parity regulations exhibit variability across states due to the diverse ways these laws are formulated and expressed (*Panel (a), Figure I*). Framing of these laws can influence not only perceptions but also the actual and expected financial incentives and cost structures for both providers and patients.

The unique legal requirements for Payment Parity within state laws can be classified as types of Price Controls (such as Price Floor, Price Ceiling, or Price Parity), types of (consumer) Cost Controls (such as Cost Ceiling or Cost Parity), a combination of both, or none at all. A state may regulate that the physician reimbursement for telehealth "cannot exceed" that of in-person services, there by establishing a "Price Ceiling"; it may require that reimbursement is "at least as much as" that for in-person services, thereby creating a "Price Floor"; or it may mandate an "equal rate" for both telehealth and in-person services, thus setting "Price Parity". Similarly, a state may demand that "deductibles, co-pays, and insurance" for telehealth "cannot exceed" those for face-to-face services, leading to a "Cost Ceiling", or that they be at the "same rate" as in-person services, establishing "Cost Parity". If a state does not specify these details, it would simply be considered to have a Payment Parity law. Each type of Price or Cost Control is mutually exclusive, meaning a state can only have one: a Price Ceiling, Price Floor, or Price Parity. Similarly, a state may specify either a Cost Ceiling or Cost Parity, but not both. The varied state-wise framing generates quasi-experimental design, which has heterogeneous effects by policy type. The paper estimates the partial equilibrium causal estimates for outcomes such as Medicare Enrollment for aged and disabled, Medicare Costs, Hospital Outpatient Visits and Preventable Hospital Stays.

---

[1]Telehealth incorporates telemedicine, which is a bilateral, interactive health communications with clinicians on both ends of the exchange (e.g. videoconferenced grand rounds, x-rays transmitted between radiologists or consultations where a remote practitioner presents a patient to a specialist).

[2]These laws could have significant implications for healthcare delivery since telehealth is crucial in reducing disease exposure during health crises, benefits patients with mobility issues or chronic conditions in Health Professional Shortage Areas (HPSAs), and can achieve outcomes similar to in-person care with cost savings (Shaver (2022); Glassman, Helgeson and Kattlove (2012)).

[3]A.D.IV shows the increasing trend in the health related use of internet. A.D.III show increased county level broadband penetration owing to Federal and State efforts.

Remote healthcare delivery and access to health information are significantly enhanced by internet connectivity. Broadband rollout has been linked to economic benefits (Canzian, Poy and Schüller (2019); Chen, Ma and Orazem (2023); Haller and Lyons (2018)) and health benefits (Van Parys and Brown (2024); Tomer et al. (2020)). Previous studies often use proxies for broadband penetration such as the number of internet providers or broadband infrastructure. This study refines this metric by using granular county-level residential connection data.

This paper contributes to the expanding body of literature that compares telehealth and in-person care, and examines the implications of telehealth parity laws on healthcare utilization (Portnoy, Waller and Elliott (2020); Bavafa, Hitt and Terwiesch (2018); Ashwood et al. (2017); Cakici and Mills (2022); Reed et al. (2021); Phillips et al. (2023)). While existing studies have indicated an increase in follow-up visits post-telehealth consultations and a potential for excessive in-person visits, many are limited by small sample sizes or lack county-level analysis, thereby limiting their external validity. This study seeks to address these limitations by providing a comprehensive county-level analysis, offering more robust and reliable insights. Furthermore, previous research on the financial implications of telehealth and telehealth parity laws (Neil Ravitz (2021); Dong (2022); Cornaggia, Li and Ye (2021); Restrepo (2018)) often does not fully consider the specific framing of these laws and lacks detailed county-level analysis. This study seeks to address these gaps by investigating the distinct effects of TPLs, considering their legislative framing, and utilizing granular county-level data. This approach enables a thorough county-level economic analysis, especially of the effects on outpatient healthcare utilization. This study, thus, provides a basis our understanding the impacts of TPL on the supply of healthcare services, proxied by county-level count of physicians as initially explored in Akimitsu (2024).

The study is also related to the literature that studies the expansion of telehealth and its impact. For instance, Saharkhiz et al. (2024) study the impact of telehealth expansion on the emergency department visits, clinician encounters and cost of care per beneficiary. However, the regulatory environment created by telehealth parity laws could be a significant driving factor behind the effects shown. Moreover, the deployment of telehealth depends on internet connection, which is not accounted for in Saharkhiz et al. (2024). Tilhou, Jain and DeLeire (2024) do take the internet into account but arrive at a rather strong conclusion that telehealth expansion and high-speed internet donot increase primary healthcare utilization.[4] Further, these studies often don't consider the licensing environment, which is a significant determinant of physicians' ability to practice. The state licensure requirements and Intertate Medical Licensure Compact (IMLC) influence the physical location, remote practice capabilities of physicians and telehealth accessibility.[5] The license agreements or 'compacts', offer a more efficient route to practicing telehealth across multiple states. This study incorporates the licensure landscape explicitly into the empirical analysis.

This study contributes empirically by modeling the adoption of TPLs as a treatment, the heterogeneous state-level framing of these laws that generate distinct types of Price Controls and Cost Controls or their combinations as treatment types or policy types, and the impacts of these controls in the form of equilibrium quantity shifts as treatment effects. These estimated average treatment effects on the treated (ATT) are partial equilibrium causal estimates. The methodology represents a significant novelty in policy analysis, adopting Poisson Pseudo-Maximum Likelihood (PPML) estimator (Santos Silva and Tenreyro (2006); Gourieroux, Monfort and Trognon (1984)) in a non-linear difference-in-differences (DID) framework with staggered intervention to account for the discrete nature of count data (Chen and Roth (2024); Wooldridge (1999); Wooldridge (2023)).

This paper presents several novel contributions. To the best our knowledge, this is the inaugural paper, alongside Akimitsu (2024), to examine the economic impacts of both "Price Controls" and consumer "Cost Controls" generated by TPLs. Thus, this is the initial study to extensively discuss the implications of state-level framing of TPLs. Consequently, the paper addresses a significant "empirical gap" in the literature concerning price regulations and telehealth. Furthermore, among the studies to empirically investigate Price Controls, this is the first one to account for the impact of technology while studying the impacts of such controls—in this instance, broadband.

Price Ceilings, often discussed in relation to rent control, typically lead to shortages because they cause a reduction in the quantity supplied by producers.[6] Conversely, Price Floors, frequently examined within the context of minimum wage laws, can result in surpluses by motivating suppliers to offer more than the equilibrium quantity. However, studies on Prices Controls on the good and services markets are rare. This is one of the very few studies which empirically investigates the effect of Price Floor in the healthcare market.[7] As per conventional theories, implementing Price Controls such as Price Ceiling or Price Floor often leads to market distortions and inefficient allocation. However, the empirical literature on Price Controls, such as rent control or minimum wage, generally lacks widespread applicability and rarely considers the geographical heterogeneity of such regulations' effects. Metro and non-metro areas, or urban and rural areas, may experience different impacts. By evaluating the effects of TPLs separately for metro and non-metro areas, this paper fills a vital "spatial" gap in the economics of regulation.

---

[4]This study offers a number of methodological improvements over Tilhou, Jain and DeLeire (2024) (henceforth, TJD). TJD use the presence of providers and median speeds as proxy for internet use. The mere presence of providers and speeds bands doesn't result in actual utilization. This study uses a more accurate measure of internet penetration and utilization, that is, the residential connection data. Second, TJD use provider specialty codes cross-tabulated with the provider rendering taxonomy, which is a highly unreliable method to identify primary care visit. In fact, (Romaire, 2020) shows that most people who use outpatient services utilize primary care. Third, TJD do not look at the role of insurance and covered hospitals or specialties. The extent of coverage puts restrictions on choice and usage. Fourth, while looking at the count of primary care visits, TJD do not use count data methods when they deal with count data, which makes their estimates biased. This study makes appropriate use of count data methods in difference-in-difference settings. Fifth, TJD completely overlook the regulatory environment, which is, in fact, the main focus of this paper. Lastly, TJD looks at data only from Wisconsin, and thus, lacks external validity. This study performs a cross-country county-level analysis and overcomes this limitation.

[5]According to the Federation of State Medical Boards Telemedicine Overview (2015), 80% of states require out-of-state clinicians offering telehealth to be licensed in the patient's residing state.

A distinctive feature of TPLs is that the market equilibrium reimbursement rate for in-person service ($MERR - I$), can function as either a Price Ceiling or a Price Floor (the equivalent for consumer Cost Controls being market equilibrium cost rate ($MECR - I$)), contingent upon the state's specification. Notably, $MERR - I$ is higher than the market equilibrium reimbursement rate for telehealth service ($MERR - T$), whereas the equilibrium consumer cost for in-person ($MECR - I$) surpasses the market equilibrium cost rate for telehealth ($MECR - T$). Therefore, when $MERR - I$ serves as a Price Ceiling on $MERR - T$, it is above $MERR - T$. The effect of such a Price Ceiling can mirror or contrast that of a Price Floor depending on whether the Price Control type is binding and on the magnitude of the distance between the unregulated equilibrium price and the regulated price. Thus, the paper also seeks to bridge the "conceptual" gap in literature concerning Price Controls, as there is a lack of consensus on the effects of such regulations. Crucially, this paper acknowledges the disparity between posted prices and consumer-borne costs within the healthcare context, due to third-party insurers. Thus, it provides further clarity on the critical "theoretical" gaps in the Price Control literature, as addressed in Akimitsu (2024), by resolving discrepancies between conventional Price Control models and empirical outcomes.

Dills (2021) presents a complex narrative about the impact of price controls in the U.S. healthcare markets due to the prevalent use of third-party payments as well as lack of research consensus about the financial implications. Despite this intricacy, the disparity between healthcare providers advocating for equal pay mandates and health insurers contesting them. This dichotomy suggests that health insurers may be gaining an economic surplus. With appropriate mandates, there is a possibility for this surplus to be redistributed to providers who offer telehealth services. The study sheds light on some of these questions in the context of Medicare. Insurer adjustments can reciprocally influence both consumer costs and provider reimbursements. A regulatory change in reimbursement rates may lead to alterations in consumer-facing expenses, such as premiums and cost-sharing. Conversely, shifts in consumer cost-sharing can impact providers' financial returns. These adjustments can take various forms, such as modifications to premiums, deductibles, and co-pays, as well as changes in coverage scope and the restructuring of provider rate schedules, which may not align with policy rates. Additionally, insurers might recalibrate service networks or negotiate provider rates to manage costs within their risk pools. Although insurers act as intermediaries, their strategic responses to market and policy shifts still generate price signals, influencing consumer and provider behavior similarly to traditional market interactions.

---

[6]For further reading, refer to Glaeser and Luttmer (2003) and Bulow and Klemperer (2012). Textbook explanations can be found in Mills and Hamilton (1994), Boyes and Melvin (2010), and Taylor and Weerapana (2010).

[7]For research on Price Floors in the labor market, see Card and Krueger (1995), Deere, Murphy and Welch (1995), and Lee and Saez (2012). Examples of studies on Price Floors in goods and services markets are Hernández and Cantillo-Cleves (2024) and Griffith, O'Connell and Smith (2022), and in healthcare market is Akimitsu (2024).

Thus, the core principles of economic models that link the price received by providers to the price paid by consumers still retain some applicability, as they incorporate the tripartite relationship. This analysis focuses on these direct expenditures without including the value of consumer inputs. However, as telehealth and in-person services involve elements managed by both consumers and providers, the full price ($P$) can also be analyzed through the value produced using these inputs.[8] The Price Control would rotate the supply curves, while the Cost Controls would rotate the demand curves. Changes in physician reimbursements might invoke physician response and affect physician supply. Cost Controls could, thus, change the consumer demand or utilization, which could be manifested in the changes in hospital outpatient visits. Any change in deductibles, copay and insurance premiums, brought about by the consumer Cost Controls, would affect the cost sharing between the consumers and insurers. This is one of the channels that could directly affect Medicare costs. Other potential channels could be changes in insurance uptake or Medicare enrollment and increased demand.[9] However, the results show that the Price and Cost Controls do not have a significant effect on Medicare enrollment. Thus, the effects on Medicare costs are driven by changes in outpatient visits and preventable hospital stays. These are demand-driven primary effects on Medicare costs, not secondary effects driven by changes in Medicare enrollment itself.

The results show that Price Ceiling-Cost Parity reduces outpatient visits and decreases Medicare costs. Price Ceiling alone also reduces Medicare costs, whereas Price Parity alone increases Medicare costs. Price Ceiling alone and the Price Floor-Cost Ceiling combination significantly reduce preventable hospital stays, while the Price Parity-Cost Parity combination increases preventable hospital stays. None of these controls have any significant effect on Medicare enrolment, indicating that the observed changes in outpatient visits, preventable hospital stays, and Medicare costs are direct effects rather than secondary changes caused by shifts in Medicare enrollment. Broadband access exhibits a statistically significant positive correlation with preventable hospital stays and a negative correlation with Medicare enrollment and costs. The Interstate Licensure Compact leads to increased Medicare costs and enrolment among the aged and disabled, addressing previously unmet demand for physician services.

---

[8]See Akimitsu (2024) for a detailed exposition

[9]The paper's focus is Medicare specifically because Medicare covers more than 18% of the total population, most of whom are people who are people aged 65 and above and the remaining are younger adults with disabilities (See Potetz, Cubanski and Neuman (2023)). It is for this demographic that telehealth is crucial due to mobility problems. This is also the demographic which is likely to be the most impacted by Cost Parity.

Due to the rapid implementation of Telehealth Parity Laws (TPLs) by states during the COVID-19 pandemic, a substantial body of research has emerged examining the financial and health-related consequences post-pandemic. However, there is a paucity of studies investigating these effects in the pre-COVID-19 context. This paper addresses this critical "temporal" gap by analyzing the impact of TPLs prior to the pandemic. It is among the first to employ a nonlinear difference-in-differences approach using count data methods to derive partial equilibrium causal estimates, thus addressing a "methodological" gap. Furthermore, this study seeks to bridge existing "literature" gaps by integrating various strands of research on telehealth, price regulations, telehealth parity laws, and health economics, providing a comprehensive evaluation of the pre-COVID implications of TPLs.

## II    EMPIRICAL FRAMEWORK

### A.    Model Specification and Data Generating Process

The empirical work involves dependent variable which could be continuous or discrete count. Variables such as logged medicare costs are continuous. Variables such as total outpatient visits, preventable hospital stays and medicare enrollment are discrete count variables. Therefore, the optimal estimation method for this model is to express it in a multiplicative form and estimate it using the Poisson Quasi Maximum Likelihood Estimator (QMLE) or Pseudo Maximum Likelihood Estimator (PMLE), which efficiently manage the zero values of $Y_{j(i)t}$ (Santos Silva and Tenreyro (2006); Gourieroux, Monfort and Trognon (1984)).[10]  Incorporating county and time fixed effects, and covariates $X_{j(i)t}$, the Difference-in-Differences (DiD) model in a generalized multiplicative form is expressed as:

---

[10]As illustrated in *Figure A.D.1* in the appendix, $Y_{j(i)t}$ represents a count variables which includes zeros. Estimating the log-linearized equation via ordinary least squares (OLS) can introduce significant biases, particularly when the true model is nonlinear in its parameters, resulting in inconsistent estimates due to heteroskedasticity. Firstly, the logarithm of zero is undefined. Secondly, even if counts are strictly positive, or transformed as $ln(1+Y_{j(i)t})$, the expected value of the log-linearized error remains dependent on the covariates, leading to biased OLS estimates. Additionally, multiplicative models estimated through nonlinear least squares (NLS) may be inefficient due to the neglect of heteroskedasticity.

FIGURE I

State-wise Telehealth Parity Laws: Framing and Staggered Adoption

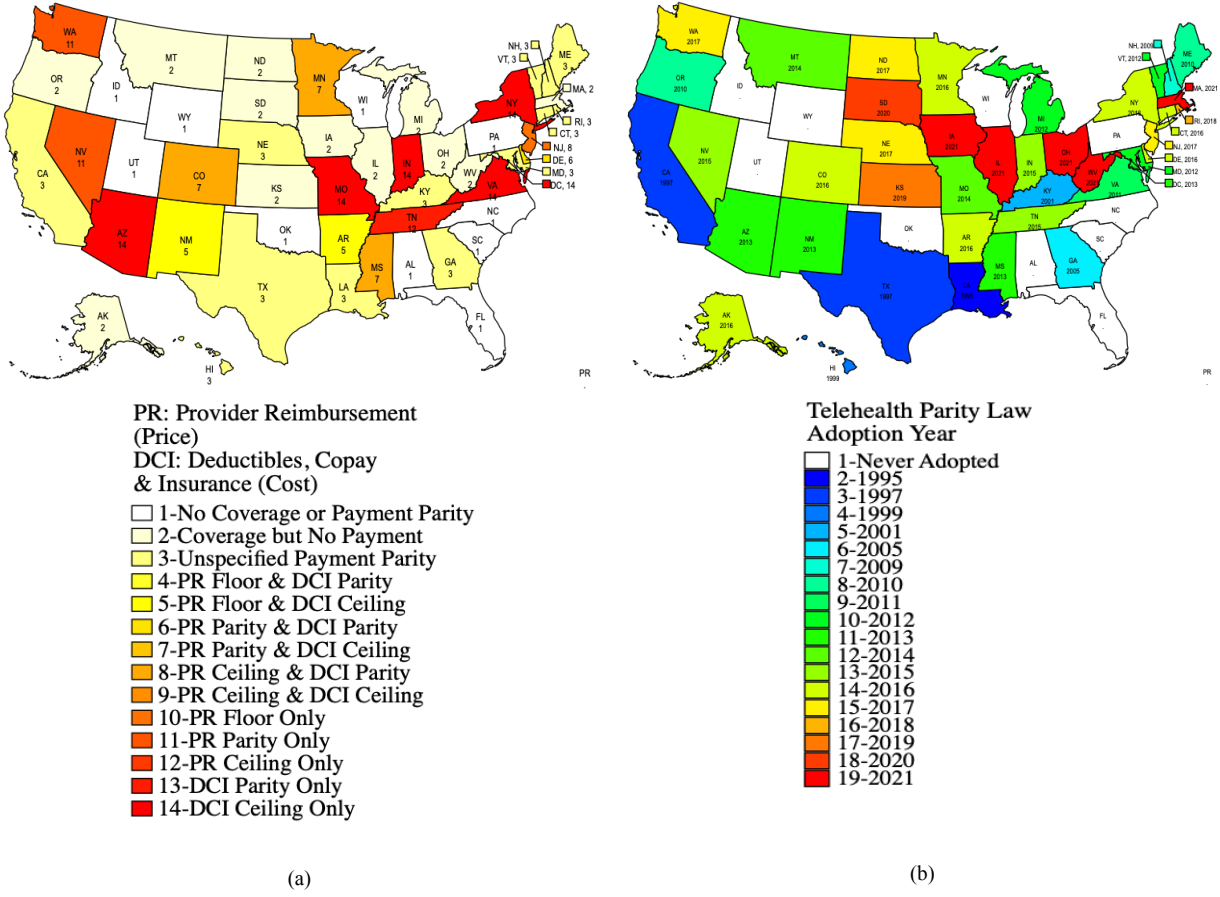

(a)

(b)

*Note*: Panel (a) shows the state level framing for the states who adopted telehealth parity laws in the United States till 2021. Panel (b) shows the staggered adoption of Telehealth Parity Laws in the United States upto 2021.

$$Y_{j(i)t} = \exp\Big( \beta_0 + \lambda_{j(i)} + \gamma_t + \sum_{k=1}^{K} \beta_{1k}(M_{ik} \times Post_{ct}) + \sum_{k=1}^{K} \beta_{2k}.M_{ik}$$

$$+ \sum_{k=1}^{K} \beta_{3k}.Post_{ct} + \beta_4 B_{j(i)t} + \beta_5' X_{j(i)t} \Big) \varepsilon_{j(i)t} \tag{1}$$

Here, $K$ is the number of treatment or regulation types considered. $c$ denotes cohort.[11] The term $\sum_{k=1}^{K} \beta_{2k}.M_{ik} + \beta_4.B_{j(i)t} + \beta_5'.X_{j(i)t}$ can be categorized into conducive, attractive, and frictional com-ponents based on the signs of the coefficients.[12] The variable $k$ denotes the treatment type as indicated by the TPL framing. Six indicators are used: one for whether the state implemented the TPL, and five for the framing types.

---

[11] Table A.B.I in Appendix, shows the cohorts and their corresponding treatment types.

Once enacted, TPLs are irreversible, precluding reversibility with staggered entry. After verifying the absence of perfect multicollinearity, the average treatment effects on the treated ($ATT_k$) for each treatment type are identified. The foundational assumptions include the "Conditional No Anticipation Assumption" and the "Conditional Indexed Parallel Trends" in levels. Recognizing that the latter assumption may be untenable, a more feasible approach is the ratio version of parallel trends. This version allows estimation of counterfactual percentage changes in mean outcomes for the treated group ($M_{1i} = 1$), based on observed changes for the never treated control group ($M_{1i} = 0$).[13] This assumption implies an exponential conditional mean function, a unique nonlinear estimation method that avoids the incidental parameters problem when including unit-specific dummies. After incorporating covariates, the Poisson PMLE, as detailed in Chen and Roth (2024) and formalized by Wooldridge (2023), consistently estimates $\theta_{\text{ATT(k)}}\%$ provided the ratio version of parallel trends holds, $\varepsilon_{j(i)t}$ has a conditional mean of 1, and $Y_{j(i)t}$ assumes an exponential form.[14] State-level main effects $M_{ik}$ are omitted as they are constant within groups and absorbed by county fixed effects $\lambda_{j(i)}$, while $Post_{ct}$ effects for each treatment type are absorbed by time fixed effects $\gamma_t$. Thus, Equation (2) simplifies to:

$$Y_{j(i)t} = \exp\left(\beta_0 + \lambda_{j(i)} + \gamma_t + \sum_{k=1}^{K} \beta_{1k}(M_{ik} \times Post_{ct}) + \beta_4 B_{j(i)t} + \beta_5' X_{j(i)t}\right)\varepsilon_{j(i)t}$$

In $M_{ik}$, $k$ will be: Price Floor, Price Ceiling, Price Parity, Cost Parity, and Cost Ceiling, when treatments are seen in isolation (Tables A.C.I and A.C.II. When types of Price Control-Cost Control are examined, as actually specified by the states, given the sample, $k$ will be: Price Floor—Cost Ceiling, Price Parity—Cost Parity, Price Parity—Cost Ceiling, Price Ceiling—Cost Parity, Price Parity only, Price Ceiling only, Cost Ceiling only (Table A.C.III). The specification includes the term $M_{i1}$, which indicates if a state adopted the TPL, corresponding to $\beta_{11}$, and the term $M_{i2}$, indicating whether the state adopted Price Parity, correspond-ing to $\beta_{12}$.

---

[12] The decision to logarithmize a covariate was based on whether it exhibited a log-normal distribution upon transformation (Beyer, Schewe and Lotze-Campen, 2022).

[13] The ratio version of the parallel trends assumption is given by:
$\dfrac{E[Y_{j(i)t}(0) \mid M_{1i} = 1, \text{Post}_{ct} = 1]}{E[Y_{j(i)t}(0) \mid M_{1i} = 1, \text{Post}_{ct} = 0]} = \dfrac{E[Y_{j(i)t}(0) \mid M_{1i} = 0, \text{Post}_{ct} = 1]}{E[Y_{j(i)t}(0) \mid M_{1i} = 0, \text{Post}_{ct} = 0]}$. If this is satisfied, the Conditional Parallel Trends Assumption is satisfied.

[14] The percentage $ATT$ for a treatment type $k$ is given by:
$\theta_{\text{ATT(k)}}\% = \dfrac{E[Y_{j(i)t}(1) \mid M_{ki} = 1, \text{Post}_{ct} = 1] - E[Y_{j(i)t}(0) \mid M_{ki} = 1, \text{Post}_{ct} = 1]}{E[Y_{j(i)t}(0) \mid M_{ki} = 1, \text{Post}_{ct} = 1]}$.

The data spans from 2010 to 2019. States treated in or before 2011 are labeled as "always treated" and excluded to avoid skewing the results, treated as missing data for those years, while states treated in 2018 or later are considered "never treated". A two-year pre-treatment period establishes baseline conditions, while a two-year post-treatment period assesses persistence or decay of treatment effects. Therefore, the earliest treated cohort includes states treated in 2012, and the latest treated cohort includes those treated in 2017.[15] While Poisson QMLE (or PMLE) difference-in-difference is employed for count data with zeros, such as Medicare enrollment, total outpatient days, or preventable hospital stays, a linear difference-in-difference model is utilized for continuous variables, such as logged Medicare costs.[16]

## III DATA

***State Level Telehealth Parity Laws Data***: The TPL information is derived from the survey report by Lacktman and Acosta (2021). The TPL languages have been tabulated by Dills (2021), detailing the adoption specifics of telehealth parity laws and their frameworks at the state level. States are grouped into cohorts based on the year they received treatment. Treatment indicators are defined by the "type" of treatment, categorizing states that received similar treatments—whether a Price Control, a Cost Control type, or a combination of both.

***County Level Broadband and Demographic Data***: This dataset comes from the Federal Communications Commission's (FCC) Form 477 County Data on Internet Access Services (Federal Communications Commission, 2024). This study is one of the first, along with to utilize this new dataset, which provides details on the number of residential broadband connections at the county level. These connections offer a precise measurement of broadband penetration. The residential connection variable is converted into a weighted standardized variable, as detailed in Section A.D.II in Appendix. This data is combined with the Staff Block Estimates (Federal Communications Commission, 2020 & 2021), which provide the county-level control variables.

***Licensure Compact Data***: The Interstate Medical Licensure Compact simplifies the licensure process for physicians practicing in multiple states. Data for states that joined the compact in 2015 and 2016 is

$$Y_{j(i)t} = \lambda_{j(i)} + \gamma_t + \sum_{k=1}^{K} \beta_{1k}(M_{ik} \times Post_{ct} \times B_{j(i)t}) + \sum_{k=1}^{K} \beta_{2k}(M_{ik} \times Post_{ct}) +$$

$$\sum_{k=1}^{K} \beta_{3k}(M_{ik} \times B_{j(i)t}) + \sum_{k=1}^{K} \beta_{4k}(Post_{ct} \times B_{j(i)t}) + B_{j(i)t} + \beta'_6 X_{j(i)t} + \varepsilon_{j(i)t}$$

obtained from Interstate Medical Licensure Compact Commission (2015-2023), available on the official website.[17]

---

[15]The estimation is performed using the procedure outlined in Correia, Guimarães and Zylkin (2020).

[16]If a broadband is interacted with Post treatment indicator, the triple interaction mdoel becomes:

**County Level Physician Count Data**: Aggregate and specialty-specific counts of physicians at the county level, along with population demographics used as control variables, are sourced from the Area Health Resource File (AHRF) (U.S. Department of Health and Human Services, Health Resources and Services Administration, Bureau of Health Workforce, 2021-2022). Released annually by the Bureau of Health Workforce, each county is uniquely identified by a "County Code" based on Federal Information Processing Standards (FIPS). The sample spans from 2010 to 2019.[18]

**Metro and Non-Metro County Classification**: Data from the 2013 Rural-Urban Continuum Codes (U.S. Department of Agriculture, Economic Research Service, 2013) is utilized to categorize U.S. counties based on urbanization, population size, and proximity to metropolitan areas. This classification aids in analyzing spatial distinctions, such as metro and non-metro areas within the U.S.[19]

## IV    RESULTS

### A.    Impact on Hospital Outpatient Visits and Medicare Costs

*Table A.C.I*  Columns 1 to 3 provide PPML DiD estimates on total hospital outpatient visits, while Columns 4 to 6 provide linear DiD estimates for the natural log of Medicare costs.[20] The Medicare costs are the costs incurred by Medicare. The significant negative coefficients for Cost Parity on hospital outpatient visits imply that Cost Parity invokes a demand-side response, leading to a reduction in outpatient visits. This is reasonable since Cost Parity is binding, and hence it increases the "deductibles, copay, and premiums" for telehealth to match those for in-person services, thereby effectively raising consumer costs for telehealth services. A Cost Ceiling, on the other hand, is not binding since the market equilibrium cost rate for telehealth (MECR-T) is already less than the market equilibrium cost rate for in-person services (MECR-I). Thus, it doesn't invoke a noticeable demand-side response. Similarly, Price Floor and Price Ceiling regulate physician reimbursement, and are unlikely to invoke demand-side responses. Broadband doesn't seem to have any impact on total hospital outpatient visits.

For the natural log of Medicare costs, the Price Ceiling, although not binding, might prevent physician reimbursement for telehealth from rising to the level of in-person reimbursement. The third-party insurer rate schedules might not allow MERR-T to rise much above its current level. As the usage of telehealth grows, the costs incurred by Medicare decrease, as shown by the strongly significant negative estimates. The non-binding Cost Ceiling also shows a negative effect on Medicare costs. However, a Cost Ceiling might not deter the insurers from passing on the costs to consumers, which would in turn reduce Medicare expenditure. On the other hand, Cost Parity is binding and raises consumer costs for telehealth usage. With both Price Parity and Price Floor scenarios, the alignment or increase in reimbursement rates for telehealth services is designed to incentivize providers to offer telehealth services and potentially increase access for patients. Consequently, more providers might engage in offering telehealth services due to favorable reimbursement rates, thus expanding the availability and use of these services. The convenience of telehealth might drive higher utilization. Greater access can be beneficial for patient outcomes but can also lead to increased overall Medicare spending if more services are being utilized (row 5, columns 4 to 6). Thus, it becomes essential to study the effects of combinations of Price Controls with Cost Controls, as actually specified by states.

---

[17]Figure A.D.V in Appendix displays the states that joined the compact in 2015 and 2016.

[18]Section A.D.I in Appendix provides additional details on the sample.

[19]Panel (a), Figure A.D.II in the Appendix, elaborates on urbanization categories.

[20]All models abstract the triple interaction involving broadband. This allows a simple comparison between counties in states which passed the TPL and counties in states which didn't pass the TPL, while controlling for broadband.

*B.  Impact on Medicare Enrollment and Preventable Hospital Stays Rate*

Table A.C.II provides insight into the impact of Telehealth Parity Laws (TPLs) on Medicare enrollment and preventable hospital stays rate. The results indicate that Cost Parity did not significantly affect Medicare enrollment, as evidenced by the coefficients in row 5, columns 1 and 3. This suggests that the observed reduction in hospital outpatient visits under Cost Parity is due to a pure demand effect rather than changes in enrollment numbers. Furthermore, Cost Parity demonstrates a strongly significant positive effect on the preventable hospital stays rate for both the aggregate sample and the metro subsample (row 5, columns 4 and 6). This implies that Cost Parity may lead to a substitution from outpatient visits to hospital stays, which could have been prevented. In metro areas, where broadband penetration is typically higher, the potential for using telehealth services is greater. However, if telehealth visits cost the same as in-person visits, consumers might prefer to pay for in-person hospital visits. Reasons for these preventable hospital stays could include delayed care due to preferences for in-person visits, utilization of emergency services during off-hours resulting in admissions, or suboptimal management of chronic conditions without the advantage of telehealth. Additionally, financial incentives for hospitals to admit patients, even for conditions that could have been managed as outpatients, might also contribute to higher hospitalization rates. Additionally, although there is a significant negative effect of a Cost Ceiling on Medicare costs in the non-metro subsample (as shown in row 6, column 2 of the Hospital Outpatient Visits table), this does not translate into a reduction in hospital outpatient visits. This difference suggests that while cost containment policies like Cost Ceilings may reduce Medicare expenditure, they do not necessarily impact the volume of outpatient services, indicating different underlying healthcare dynamics and utilization patterns between metro and non-metro areas.

*C. Impact of the combinations of Price Controls with Cost Controls*

The effect on Medicare outpatient costs could come from various channels - utilization, Medicare enrollment, or preventable hospital stays. If effect on Medicare enrollment is negligible or insignificant, it implies that the increase in Medicare outpatient costs may be due to increase in outpatient visits. If effect on outpatient visits is negligible or insignificant, then the increase in Medicare outpatient costs could be driven by increase in preventable hospital stays. The impact of combinations of types of Price Controls with those of Cost Controls creates a clearer picture.

The adoption of Price Controls or Cost Controls, or their combinations, by states necessitates an understanding of their distinct and combined impacts to identify the true effects of each. Table A.C.III elucidates the effects of different combinations of Price Controls and Cost Controls on hospital outpatient visits, medicare costs, preventable hospital stays, and Medicare enrollment.

As observed in Table A.C.I, Cost Parity alone has a significant negative effect on outpatient visits. However, when combined with Price Parity, this effect is moderated but remains negative (row 3, column 1 in Table A.C.III). Similarly, combining Cost Parity with a Price Ceiling results in an effect analogous to examining Cost Parity in isolation, indicating that the Price Ceiling's contribution is minuscule. This is corroborated by the results in row 7, column 1 in Table A.C.III and row 4 in Table A.C.I. The demand-side response observed in the reduction of outpatient visits, as discussed, cannot be attributed to a decrease in Medicare enrollment (column 4 in Table A.C.III); thus, it reflects a pure demand response.

Price Ceiling alone induces a significant negative effect on the log of Medicare costs (row 4, columns 4 to 6 in Table A.C.I and row 7, column 2, in table A.C.III). Conversely, Cost Parity exhibits a statistically significant positive effect on Medicare costs for the aggregate sample (row 5, columns 4 in Table A.C.I). Cost Parity, however, doesn't occur in isolation, i.e., in the given sample duration, no state opted for Cost Parity alone, but only in combination with other types of Price Controls. When examining combinations of Price and Cost Controls, it is evident that the Price Ceiling-Cost Parity combination yields a significant negative effect on Medicare costs, though to a lesser extent compared to the Price Ceiling alone (row 7, column 2 in Table A.C.III). The effect of Cost Parity on preventable hospital stays, whether combined with Price Parity or Price Ceiling, remains consistent (rows 3 and 5, column 3 in Table A.C.III). $MERR-T$ pre-regulation was already less than $MERR-I$. Thus, there is scope for physician reimbursement for telehealth to rise when a Price Ceiling is specified, provide the insurer rate schedules adjust accordingly. Thus, Price Ceiling also cause increased reimbursement of physicians for telehealth. This incentivises physicians to offer more telehealth visits, which reduces the preventable hospital stays rate. Row 7, column 3 in table A.C.III confirms that. Price Floor, on the other hand, is binding and must raise the physician reimbursement. Depending on the type of Cost Control it is combined with, Price Floor would also incentivize physicians to offer more telehealth visit, which would reduce the preventable hospital stays rate. The negative estimate for the coefficient in row 2, column 3 shows that this is indeed the case. A combination of Price Control with Cost Ceiling reduced the preventable hospital stays rate.

*Impact of Broadband:* Broadband shows a statistically significant positive correlation with preventable hospital stays (row 9, column 3 in table A.C.III). This positive correlation could be due to improved health awareness and easier access to healthcare information, leading patients to seek more care, including hospital stays that might have been preventable with better management. Broadband also shows a statistically significant negative correlation with Medicare enrollment for aged and disabled (Table A.C.III, column 4). This relationship suggests that higher broadband penetration may facilitate better access to information, allowing individuals to choose insurance products other than Medicare. This improved access to information could result in more informed decisions regarding insurance options. Additionally, a decrease in Medicare enrollment is naturally associated with a decrease in log Medicare costs (Table A.C.III, column 2). The greater access to information through broadband potentially enables consumers to sort into better insurance products, reducing reliance on Medicare and, subsequently, decreasing Medicare costs. Broadband appears to have a negligible correlation with hospital outpatient visits (Table A.C.III, row 9, column 1). This indicates that while broadband penetration enhances access to information and facilitates better decision-making regarding healthcare and insurance products, it does not significantly alter the consumer demand for outpatient visits. In summary, broadband is correlated with better access to health information, leading to more informed insurance choices and reduced Medicare costs, rather than directly affecting the demand for hospital outpatient visits.

*With Broadband Interaction*: *Table A.C.IV* provides the results when the posttreatment indicators are interact with the standardized broadband index. Rows 1 to 6 show the estimates for $\beta_{11}$ to $\beta_{16}$ from (3). This helps identify not only whether treatment as a whole has an effect ($\beta_{11}$ in (3)), but also whether specific types of treatment have different effects ($\beta_{12}$ to $\beta_{16}$ in (3)). The estimates show comparison is between counties in the untreated states (states which did not enact TPL) at the mean broadband level, with the counties in the treated states (states which did enact TPL), at broadband level one standard deviation above the mean. The dependent variable is the number of outpatient visits.[21] Only the estimates for Cost Parity and Price Floor show significance. Cost Parity is binding since $MECR - T$ was below $MECR - I$ pre-regulation in states which didn't enact TPL. Thus, only Cost Parity significantly invokes a demand side response for aggregate sample and metro areas. The estimate for Price Floor, even though strongly significant for the aggregate sample, is very minuscule. The strongly significant estimate for non-metro subsample, with broadband level one SD above the mean for Price Floor, could indicate duplicate visits (Cakici and Mills, 2022). This is because Price Floor is binding and physicians are getting reimbursed as much as in-person. So, there is incentive to book multiple appointments for the same condition for the same patient owing to the convenience offered by telehealth to both, if broadband is 1 SD above the mean. As opposed to that, Price Ceiling is not binding and hence it might not be able to create significant impact even in the presence of broadband. However, the higher magnitude in non-metro areas could be due to scale effect owing to the already low demand at the baseline. Thus, Cost Parity is the only type of regulation which convincingly involves a demand side response. In fact, Cost Parity reduces the demand, as signified by hospital outpatient visits, in areas with high broadband, since the telehealth becomes as expensive as in-person services.

---

[21]The interpretation in this case, for instance, would be: In column 6, row 5, the coefficient of -0.130 on the triple interaction term suggests that in a metro county within a state that enacted a Cost Ceiling policy, a one-standard deviation increase in broadband is associated with an increase in the expected physician count by a factor of 13% more than what would have been expected in the same county, at the mean broadband level, within the same state had the state not adopted the policy, holding other variables constant.

*Interstate Licensure Compact:* Participation in the Interstate Licensure Compact allows physicians to practice across state lines in compact-participating states, which has significant implications for telehealth. It facilitates the delivery of care to patients from their home state more conveniently. As shown in Table A.C.III, the Interstate Licensure Compact has no significant impact on hospital outpatient visits or preventable hospital stays. However, it does show a weakly significant increases in Medicare costs (column 2). This suggests that consumers previously unable to access medical care due to a shortage of physicians now access physician services from other states, meeting previously unmet demand and leading to increased Medicare costs.[22] Furthermore, the Interstate Licensure Compact shows a strongly significant positive impact on Medicare enrollment for the aged and disabled (column 4). This implies that the opportunity to meet unmet demand by employing physician services from other states might have encouraged aged and disabled consumers to enroll in Medicare to take advantage of this increased access to care. In summary, while the Interstate Licensure Compact does not significantly impact hospital outpatient visits or preventable hospital stays, it leads to increased Medicare costs and higher Medicare enrollment among the aged and disabled, likely due to enhanced access to physician services from compact-participating states.

*Possible Mechanisms:* As discussed in Akimitsu (2024), in the healthcare market, the traditional direct price-consumer interaction is replaced by a tripartite relationship involving the consumer, provider, and a third-party insurer. The full price ($P$) that consumers face includes not only direct expenditures like out-of-pocket costs but also indirect costs through insurance premiums ($S$), which are divided into the provider's share ($S_p$) and the insurer's share ($S_n$), along with administrative costs ($A$). Therefore, the total price consumers encounter is a combination of out-of-pocket expenditure ($E_{\text{oop}}$) and insurance premium components spread over the number of healthcare service units ($Y$), represented by:

---

[22]For instance, Wisconsin was able to address its physician shortage problem by adopting the interstate medical licensing legislation. The news can be accessed here: https://imlcc.com/news/press-releases-and-publications/

$$P = E_{\text{oop}} + \frac{S_p + S_n + A}{Y}$$

Here, $E_{\text{oop}} = \frac{D}{Y} + Co$, where $D$ is the annual deductible and $Co$ is the copayment per healthcare service. $\frac{S_p}{Y}$ represents the portion of the insurance premium allocated toward provider reimbursements, $\frac{S_n}{Y}$ is the insurer's profit share, and $\frac{A}{Y}$ is the allocation for the insurer's administrative costs. The regulatory policy parameters will be $\gamma$ and $\rho$ such that $P = \gamma$ and $\frac{S_p}{Y} = \rho$.

The model examines how Price Controls (affecting provider reimbursement, $\rho$) and Cost Controls (affecting consumer out-of-pocket costs, $E_{\text{oop}}$) influence the full price ($P$), which drives demand for outpatient visits ($Y$) and Medicare costs ($C$).

- **Price Parity (PP)**: Sets $\rho = MERR - I$, increasing $\rho$ since $MERR - T < MERR - I$ pre-regulation.

- **Price Ceiling (PC)**: Caps $\rho \leq \rho_{\text{max}}$, where $\rho_{\text{max}} < MERR - I$, but $\rho$ may rise if $MERR - T < \rho_{\text{max}}$.

Consumer Out-of-Pocket Costs ($E_{\text{oop}} = MECR - T$):

- **Cost Parity (CP)**: Sets $\gamma = MECR - I$, increasing $E_{\text{oop}}$ when $MECR - T < MECR - I$.

- **Cost Ceiling (CC)**: Caps $\gamma \leq MECR - I$. If $MECR - T < MECR - I$, $E_{\text{oop}}$ can rise up if insurer passes on the costs to the consumer (C decreases), making the ceiling non-binding unless $MECR - T \geq MECR - I$ or difference between pre-regulation $MECR - I$ and $MECR - T$ is $\approx 0$.

The full price ($P$) paid by the consumer is given by:

$$P = E_{\text{oop}} + \rho$$

The demand for outpatient visits ($Y$) is represented by the linear relationship:

$$Y = a - bP$$

Medicare costs ($C$) are calculated as:

$$C = \rho Y$$

The results for outpatient visits at a 1% significance level indicate a significant effect post-Price Ceiling-Cost Parity, with a coefficient of -0.156 (p < 0.01). The explanation for this involves the interplay between Price Ceiling and Cost Parity mechanisms. Price Ceiling caps $\rho \leq \rho_{\text{max}}$. Given that $MERR - T < MERR - I$ pre-regulation, $\rho$ can rise up to $\rho_{\text{max}}$, though less than under Price Parity. Cost Parity sets $E_{\text{oop}} = MECR - I$. Since $MECR - T < MECR - I$, $E_{\text{oop}}$ increases significantly. The combined effect of rising $E_{\text{oop}}$ (dominant effect) and the moderate increase in $\rho$ elevates the full price $P$. Consequently, the higher price ($P$) reduces the demand for outpatient visits, as evidenced by the substantial negative coefficient. This outcome corroborates the model's prediction that cost parity exerts a strong negative influence on visit frequency, with a price ceiling contributing modestly through $\rho$. The observed decrease in demand for outpatient visits is a straightforward demand response, independent of any changes in Medicare enrollment, the effects of which are found to be statistically insignificant.

As shown in table A.C.III outpatient Medicare costs at a 1% significance level, significant results are observed post-Price Parity only (0.079), post-Price Ceiling only (-0.145), and post-Price Ceiling-Cost Parity (-0.088). Post-Price Parity only, setting $\rho = MERR - I$ increases $\rho$ since $MERR - T < MERR - I$. Without cost control, $E_{\text{oop}}$ remains at pre-regulation levels. The higher $\rho$ increases $P$, reducing $Y$. Despite the lower $Y$, the significant rise in $\rho$ outweighs the visit reduction, resulting in increased $C = \rho Y$, particularly if demand is inelastic. This aligns with the positive coefficient (0.079), reflecting higher Medicare costs due to elevated reimbursement rates, consistent with the model's emphasis on Price Controls' cost-increasing potential when not accompanied by Cost Controls.

In the case of post-Price Ceiling only, capping $\rho \leq \rho_{\text{max}}$, $\rho$ rises but is limited compared to Price Parity. Without cost control, $E_{\text{oop}}$ remains unchanged. The modest increase in $P$ due to the constrained $\rho$ rise causes a small reduction in $Y$. Consequently, the limited $\rho$ growth restricts $C$, and lower $Y$ reinforces this. The large negative coefficient (-0.145) signifies a significant cost reduction, which aligns with the results section, noting that Price Ceiling alone induces a prominent negative effect on Medicare costs.

Post-Price Ceiling-Cost Parity combines the effects of both mechanisms. Price Ceiling limits $\rho \leq \rho_{\text{max}}$, allowing a moderate increase from $MERR - T$. Cost Parity sets $E_{\text{oop}} = MECR - I$, increasing $E_{\text{oop}}$ since $MECR - T < MECR - I$. The increases in both $E_{\text{oop}}$ and $\rho$ raise $P$, reducing $Y$ as pronounced in the -0.156 result. Consequently, lower $Y$ and constrained $\rho$ reduce $C$. The negative coefficient (-0.088) indicates a significant but smaller cost decrease compared to the Price Ceiling alone (-0.145), supporting the observation that Cost Parity moderates the Price Ceiling's negative effect on costs.

A significant result is also found for post-Cost Ceiling only at the 5% significance level, with a coefficient of -0.054 (p < 0.05). Under this mechanism, the Cost Ceiling caps $E_{\text{oop}} \leq MECR - I$. Pre-regulation, $MECR - T < MECR - I$, making the ceiling non-binding in preventing a rise up to $MECR - I$ (though it prevents exceeding it). Insurers might adjust $E_{\text{oop}}$ upward within this limit to offset costs or align with market dynamics, increasing $E_{\text{oop}}$ from $MECR - T$. In the absence of price control, $\rho$ remains at pre-regulation levels ($MERR - T$). The potential rise in $E_{\text{oop}}$ would increase $P$, assuming insurers exploit the ceiling's flexibility, though the effect is smaller than with Cost Parity's mandated increase. Higher $P$ reduces $Y$, and the negative coefficient (-0.054) reflects a moderate but significant decrease in visits, consistent with a demand response to a flexible, non-binding ceiling. This aligns with the results section's emphasis on demand-side effects independent of enrollment changes.

Additional insights from the results section suggest that Cost Parity consistently reduces outpatient visits (e.g., via higher $E_{oop}$), but its effect is moderated when paired with Price Controls. This supports the model's prediction that increased $P$ drives down $Y$. Price Ceiling's impact limits $\rho$ growth, significantly reducing Medicare costs (e.g., -0.145 alone), though less so with Cost Parity (-0.088), reflecting the interplay of $Y$ and $\rho$. Furthermore, the reduction in visits is not attributed to Medicare enrollment drops, reinforcing the model's demand-driven mechanism. Additionally, broadband correlates positively with preventable hospital stays and negatively with Medicare enrollment, but its negligible effect on outpatient visits suggests it influences insurance choices rather than visit demand, supporting the model's focus on $P$.

Certain Telehealth Payment Parity Law (TPL) combinations significantly reduce Preventable Hospital Stays. For instance, the Price Floor-Cost Ceiling combination leads to a 5.3% decrease, while the Price Ceiling alone results in a 6.6% decrease. These reductions arise because TPLs increase provider reimbursements and adjust consumer costs, encouraging greater use of outpatient telehealth visits. This shift enhances access to preventive care, effectively lowering the incidence of preventable hospitalizations. In contrast, Medicare Enrollment for Aged and Disabled remains unaffected by TPLs, with statistically insignificant changes such as a 0.9% decrease under the Price Floor-Cost Ceiling combination. Enrollment decisions are instead driven by factors like eligibility criteria and premium costs, not TPL-induced changes. Consequently, variations in Medicare Costs—such as a 7.9% increase under Price Parity or an 8.8% decrease under Price Ceiling-Cost Parity—are not due to enrollment fluctuations but reflect changes in utilization, as demonstrated by Outpatient Visits (e.g., a 15.6% decrease under Price Ceiling-Cost Parity). These utilization shifts result from demand-side responses to the altered full price of care under TPLs, particularly through Cost Controls, which directly influence consumer costs and thus the demand for outpatient telehealth services.

## V ROBUSTNESS CHECKS

To ensure balance and comparability of treatment and control groups, logistic regression-based Propensity Score Matching (PSM) with multiple explanatory variables and comprehensive model diagnostics were performed. Multicollinearity was evaluated and sensitivity analyses were performed. After computing propensity scores, their distribution by treatment status was checked via histograms, and matching was implemented with a caliper width of 0.05 to create comparable treated and control groups. Balance diagnostics post-matching confirmed the comparability of these groups, with all variables showing biases below the 25% threshold. To ensure the correctness of the functional form while using the Fixed Effects Poisson Estimator, the heteroskedasticity-robust Ramsey's Regression Specification Error Test (RESET) was applied (Ramsey, 1969). County and year fixed effects were included to control for unobserved, time-invariant characteristics and common time trends. Table A.E.I shows that the test showed that there are no signs of misspecification.[23] A placebo test for pre-treatment trends was conducted by assigning a fictitious treatment period. Table A.E.II test showed no significant pre-treatment differences between treated and control groups, suggesting that the treatment effect estimates are not driven by pre-existing trends or other exogenous factors. The ratio version of the parallel trends assumption was tested by conducting a PPML event study, where "relative time from treatment" was interacted with treatment indicator. The results support the assumption (Table A.E.III). To test the no-anticipation assumption, PPML model with broadband interaction on the lag of count of MDs was estimated. The results support the assumption that the future treatment does not affect current outcomes (Table A.E.IV).

## VI    CONCLUSION

This study addresses an important gap in the literature by analyzing the impact of Telehealth Parity Laws on outpatient visits, Medicare enrollment, preventable hospital stays, and Medicare costs in the pre-pandemic era. The staggered adoption and heterogeneity in the framing of these regulations have seldom been taken into account. The study finds that the effect on these four outcome variables at the county level differs according to the framing by each state. The framing differs due to different stipulations by each state regarding physician reimbursement and deductibles, copay, and insurance premium paid by the consumers for telehealth as compared to those for in-person services. This framing by each state generates various types of Price Controls, often combined with types of consumer Cost Controls. The conventional literature on price regulations predicts shortages or inefficiencies due to regulations. The study shows that such assertions are not necessarily true. The actual impact of a type of Price Control or Cost Control depends on whether it is used in isolation or combination, as well as on the geographical area—whether it is metro or non-metro.

By studying the Cost Controls in conjunction with Price Controls, the study finds that Cost Parity, which effectively increases the out-of-pocket expenditure of the consumers for telehealth, has a negative effect on the outpatient visits when combined with either of the Price Controls. Price Parity effectively increases the physician reimbursement for telehealth to that of in-person. Price Ceiling prevents the physician reimbursement for telehealth from exceeding that of in-person. Thus, Price Parity is more favorable from the physician's point of view. A Price Ceiling-Cost Parity combination has a strongly significant negative impact on hospital outpatient visits. Price Parity-Cost Parity also has a negative effect on the outpatient visits. However, the effect is much moderate than that of Price Ceiling-Cost Parity combination, since Price

---

[23]Some of the estimates show higher standard errors, particularly for the non-metro subsample. Non-metro counties differ significantly in their infrastructure leading to greater inherent variability or instability in responses to policy changes in these areas. Additionally, the complexity of the model further contributes to this issue. Nevertheless, the use of cluster-robust standard errors at the state level corrects for both heteroskedasticity and autocorrelation within counties, and accounts for intragroup correlation, resulting in more accurate standard errors and reliable estimates.

Parity is favorable for the physician and incentivizes more outpatient visits.

In terms of Medicare costs, the study finds that Price Ceiling alone significantly reduces the log of Medicare costs, while Cost Parity tends to increase these costs when assessed independently. Combinations involving Price Ceiling and Cost Parity still lead to overall reductions in Medicare costs, but the extent of the reduction is moderated compared to the Price Ceiling alone. Regarding preventable hospital stays, the Price Parity-Cost Parity combination results in a significant increase, indicating reduced efficiency in patient care management. The Price Ceiling-Cost Parity combination also increases preventable hospital stays, but the effect is less pronounced compared to the Price Parity-Cost Parity combination. The Price Ceiling used alone significantly reduces preventable hospital stays. Additionally, the Price Floor-Cost Ceiling combination significantly reduces preventable hospital stays, indicating that this combination effectively reduces unnecessary admissions.

Broadband access is associated with reduced Medicare costs and Medicare enrollment, likely due to better access to alternative insurance information. However, broadband shows a positive correlation with preventable hospital stays, suggesting that better internet access might lead to more hospital stays that could have been prevented. Broadband does not significantly affect hospital outpatient visits. The Interstate Licensure Compact shows no significant impact on hospital outpatient visits or preventable hospital stays. It is associated with a weakly significant increase in Medicare costs, likely due to addressing previously unmet demand for physician services. It also significantly increases Medicare enrollment among aged and disabled individuals, enhancing access to care across state lines.

In conclusion, the detailed impacts of Telehealth Parity Laws underscore the importance of carefully considering the design and combination of Price and Cost Controls. Policymakers should account for specific contexts and stakeholder incentives to fully leverage telehealth's potential in improving healthcare access, reducing preventable hospitalizations, and managing healthcare costs effectively. Broadband access and the Interstate Licensure Compact further emphasize the need for comprehensive telehealth policy formulation.

## REFERENCES

**Akimitsu, Piyush.** 2024. "Price Regulation, Technology and Provider Redistribution." 2, 3, 4, 5, 14

**Ashwood, J. Scott, Ateev Mehrotra, Daniel Cowling, and Lori Uscher-Pines.** 2017. "Direct-To-Consumer Telehealth May Increase Access To Care But Does Not Decrease Spending." *Health Affairs (Project Hope)*, 36(3): 485–491. 2

**Bavafa, Hessam, Lorin M. Hitt, and Christian Terwiesch.** 2018. "The Impact of E-Visits on Visit Frequencies and Patient Health: Evidence from Primary Care." *Management Science*, 64(12): pp. 5461–5480. 2

**Beyer, R. M., J. Schewe, and H. Lotze-Campen.** 2022. "Gravity models do not explain, and cannot predict,

international migration dynamics." *Humanit Soc Sci Commun*, 9: 56. 8

**Blewett, Lynn A., Julia A. Rivera Drew, Miriam L. King, Kari C.W. Williams, Annie Chen, Stephanie Richards, and Michael Westberry.** 2023. "IPUMS Health Surveys: National Health Interview Survey, Version 7.3 [dataset]." *Minneapolis, MN: IPUMS*, DOI: https://doi.org/10.18128/D070.V7.3. 38

**Boyes, William, and Michael Melvin.** 2010. *Microeconomics.* . 8th ed., Mason, OH:Cengage. 4

**Bulow, Jeremy, and Paul Klemperer.** 2012. "Regulated Prices, Rent-Seeking, and Consumer Surplus." *Journal of Political Economy*, 120(1): 160–186. 4

**Cakici, Ozden, and Alex Mills.** 2022. "Telehealth in Acute Care: Pay Parity and Patient Access." SSRN. 2, 14

**Canzian, Giulia, Samuele Poy, and Simone Schüller.** 2019. "Broadband upgrade and firm performance in rural areas: Quasi-experimental evidence." *Regional Science and Urban Economics*, 77: 87–103. 2

**Card, David, and Alan B. Krueger.** 1995. "Time-Series Minimum-Wage Studies: A Meta-analysis." *The American Economic Review*, 85(2): 238–243. 4

**Chen, Jiafeng, and Jonathan Roth.** 2024. "Logs with Zeros? Some Problems and Solutions." *The Quarterly Journal of Economics*, 139(2): 891–936. 3, 8, 31, 32

**Chen, Yulong, Liyuan Ma, and Peter F. Orazem.** 2023. "The heterogeneous role of broadband access on establishment entry and exit by sector and urban and rural markets." *Telecommunications Policy*, 47(3): 102504. 2

**Cornaggia, Kimberly Rodgers, Xuelin Li, and Zihan Ye.** 2021. "Virtual Competition and Cost of Capital: Evidence from Telehealth." *HEN: Public Health (Topic)*. 2

**Correia, Sergio, Paulo Guimarães, and Tom Zylkin.** 2020. "Fast Poisson estimation with high-dimensional fixed effects." *The Stata Journal*, 20(1): 95–115. 9

**Deere, Donald, Kevin M. Murphy, and Finis Welch.** 1995. "Employment and the 1990-1991 Minimum-Wage Hike." *The American Economic Review*, 85(2): 232–237. 4

**Dills, Angela K.** 2021. "The Effect of Price Minimums in U.S Healthcare Markets." Mercatus Center, George Mason University Policy Brief. Available at: https://www.mercatus.org/media/74946/download?attachment. 4, 9

**Dong, Xiaoyu.** 2022. "The Impact of Telehealth Parity Laws on Health Expenses." SSRN. 2

**Federal Communications Commission.** 2020 & 2021. "Staff Block Estimates [data]." *Economics and Analytics*, Available at: https://www.fcc.gov/staff-block-estimates. 9

**Federal Communications Commission.** 2024. "Historical Form 477 County-Level Tier Data [data]." Available at: https://www.fcc.gov/form-477-county-data-internet-access-services. 9

**Glaeser, Edward L., and Erzo F. P. Luttmer.** 2003. "The Misallocation of Housing under Rent Control."

*The American Economic Review*, 93(4): 1027–1046. 4

**Glassman, Paul, Michael Helgeson, and Jean Kattlove.** 2012. "Using telehealth technologies to improve oral health for vulnerable and underserved populations." *Journal of the California Dental Association*, 40(7): 579–585. 1

**Gourieroux, C., A. Monfort, and A. Trognon.** 1984. "Pseudo Maximum Likelihood Methods: Theory." *Econometrica*, 52(3): 681–700. 3, 6

**Griffith, Rachel, Martin O'Connell, and Kate Smith.** 2022. "Price Floors and Externality Correction*." *The Economic Journal*, 132(646): 2273–2289. 4

**Haller, Stefanie, and Sean Lyons.** 2018. "Effects of broadband availability on total factor productivity in service sector firms: Evidence from Ireland." *Telecommunications Policy*, 43. 2

**Hernández, Carlos Eduardo, and Santiago Cantillo-Cleves.** 2024. "A toolkit for setting and evaluating price floors." *Journal of Public Economics*, 232: 105084. 4

**Interstate Medical Licensure Compact Commission.** 2015-2023. "Participating States: 2015 and 2016 [data]." *https://imlcc.com/participating-states/*. 10

**Lacktman, Nathaniel M., and Jacqueline N Acosta.** 2021. "50-State Survey of Telehealth Commercial Insurance Laws." *Foley and Lardner LLP*, Available at: https://www.foley.com/insights/publications/2021/02/50-state-survey-of-telehealth-insurance-laws/. 9

**Lee, David, and Emmanuel Saez.** 2012. "Optimal Minimum Wage Policy in Competitive Labor Markets." *Journal of Public Economics*, 96: 739–749. 4

**Mills, Edwin S., and Bruce W. Hamilton.** 1994. *Urban Economics.* New York, N.Y.:HarperCollins. 4

**Naqvi, Asjad.** 2023. "Stata package "bimap"." Available at: https://github.com/asjadnaqvi/stata-bimap. 36

**Neil Ravitz, M. B. A.** 2021. "The Economics of a Telehealth Visit: A Time-Based Study at Penn Medicine." *https://www.hfma.org/topics/financial-sustainability/article/the-economics-of-a-telehealth-visit--a-time-based-study-at-penn-.html*. 2

**Phillips, J. C., R. W. Lord, S. W. Davis, A. A. Burton, and J. K. Kirk.** 2023. "Comparing Telehealth to Traditional Office Visits for Patient Management in the COVID-19 Pandemic: A Cross-sectional Study in a Respiratory Assessment Clinic." *Journal of Telemedicine and Telecare*, 29(5): 374–381. 2

**Portnoy, Jay, Mark Waller, and Timothy Elliott.** 2020. "Telemedicine in the Era of COVID-19." *The Journal of Allergy and Clinical Immunology: In Practice*, 8(5): 1489–1491. 2

**Potetz, Lisa, Juliette Cubanski, and Tricia Neuman.** 2023. "Medicare Spending and Financing." Health Policy Alternatives, Inc. and The Henry J. Kaiser Family Foundation. Prepared by: Lisa Potetz, Health Policy Alternatives, Inc. and Juliette Cubanski and Tricia Neuman, The Henry J. Kaiser Family Foundation. 5

**Ramsey, J. B.** 1969. "Tests for Specification Errors in Classical Linear Least-Squares Regression Analysis." *Journal of the Royal Statistical Society. Series B (Methodological)*, 31(2): 350–371. 17

**Reed, Mary, Jie Huang, Ilana Graetz, Emilie Muelly, Andrea Millman, and Catherine Lee.** 2021. "Treatment and Follow-up Care Associated With Patient-Scheduled Primary Care Telemedicine and In-Person Visits in a Large Integrated Health System." *JAMA Network Open*, 4(11): e2132793–e2132793. 2

**Restrepo, K.** 2018. *The Case against Telemedicine Parity Laws.* John Locke Foundation, Raleigh, NC. 2

**Romaire, Melissa A.** 2020. "Use of Primary Care and Specialty Providers: Findings from the Medical Expenditure Panel Survey." *Journal of General Internal Medicine*, 35(7): 2003–2009. 3

**Saharkhiz, Morteza, Tanvi Rao, Sara Parker-Lue, Sara Borelli, Karin Johnson, and Guido Cataife.** 2024. "Telehealth Expansion and Medicare Beneficiaries' Care Quality and Access." *JAMA Network Open*, 7(5): e2411006–e2411006. 2

**Santos Silva, João, and Silvana Tenreyro.** 2006. "The Log of Gravity." *The Review of Economics and Statistics*, 88(4): 641–658. 3, 6

**Shaver, J.** 2022. "The State of Telehealth Before and After the COVID-19 Pandemic." *Prim Care*, 49(4): 517–530. Epub 2022 Apr 25. PMID: 36357058; PMCID: PMC9035352. 1

**Sun, Liyang, and Sarah Abraham.** 2021. "Estimating dynamic treatment effects in event studies with heterogeneous treatment effects." *Journal of Econometrics*, 225(2): 175–199. Themed Issue: Treatment Effect 1. 28

**Taylor, John B., and Akila Weerapana.** 2010. *Principles of Microeconomics. .* 7th ed., Mason, OH:Cengage. 4

**Tilhou, Alyssa Shell, Arjun Jain, and Thomas DeLeire.** 2024. "Telehealth Expansion, Internet Speed, and Primary Care Access Before and During COVID-19." *JAMA Network Open*, 7(1): e2347686–e2347686. 2, 3

**Tomer, Adie, Lara Fishbane, Angela Siefer, and Bill Callahan.** 2020. "Digital Prosperity: How Broadband Can Deliver Health and Equity to All Communities." *Brookings.* Accessed on: [Feb 15, 2024]. 2

**U.S. Census Bureau.** 2016. "2016 TIGER/Line Shapefiles (machine-readable data files)[data]." prepared by the U.S. Census Bureau, 2016. 36

**U.S. Department of Agriculture, Economic Research Service.** 2013. "Rural-Urban Continuum Codes [data]." Available at: https://www.ers.usda.gov/data-products/rural-urban-continuum-codes/. 10

**U.S. Department of Health and Human Services, Health Resources and Services Administration, Bureau of Health Workforce.** 2021-2022. "Area Health Resources Files (AHRF) [data]." *Rockville, MD*, Available at: https://data.hrsa.gov/data/download?data=AHRF. 10

**Van Parys, Jessica, and Zach Y. Brown.** 2024. "Broadband Internet access and health outcomes: Patient and provider responses in Medicare." *International Journal of Industrial Organization*, 95: 103072. 2

**Wooldridge, Jeffrey M.** 1999. "Distribution-free estimation of some nonlinear panel data models." *Journal of Econometrics*, 90(1): 77–97. 3

**Wooldridge, Jeffrey M.** 2023. "Simple approaches to nonlinear difference-in-differences with panel data." *The Econometrics Journal*, 26(3): C31–C66. 3, 8

APPENDIX

Summary Statistics

Table A.B.I—: Framing types (treatment types) according to cohort (year first treated).

| | Framing Type | Never Treated | 2012 | 2013 | 2014 | 2015 | 2016 | 2017 |
|---|---|---|---|---|---|---|---|---|
| (1) | **Never Treated** | 1 | 0 | 0 | 0 | 0 | 0 | 0 |
| (2) | **Price Floor-Cost Ceiling** | 0 | 0 | 1 | 0 | 0 | 1 | 0 |
| (3) | **Price Parity-Cost Parity** | 0 | 0 | 0 | 0 | 0 | 1 | 0 |
| (4) | **Price Parity-Cost Ceiling** | 0 | 0 | 1 | 0 | 0 | 1 | 0 |
| (5) | **Price Ceiling-Cost Parity** | 0 | 0 | 0 | 0 | 0 | 0 | 1 |
| (6) | **Price Parity only** | 0 | 0 | 0 | 0 | 1 | 0 | 1 |
| (7) | **Price Ceiling only** | 0 | 0 | 0 | 0 | 1 | 0 | 0 |
| (8) | **Cost Ceiling only** | 0 | 0 | 1 | 1 | 1 | 1 | 0 |
| (9) | **No Framing** | 0 | 1 | 1 | 1 | 1 | 1 | 1 |

(2) to (8) are various treatment types. (9) is when a state doesn't specify a framing type, even though it adopted the Telehealth Parity Law (TPL). Thus, the main treatment, whether a state has adopted the TPL or not is collectively (2) to (9). (1) is the never treated group.

Table A.B.II—: Summary Table

| | Never Treated (49.3%) | Treated (50.7%) | Test |
|---|---|---|---|
| Tot Hosps - Outpatient Visits | 2.7e+05 (8.6e+05) | 2.6e+05 (7.4e+05) | |
| Total Actual Medicare Costs | 1.1e+08 (2.8e+08) | 1.0e+08 (2.6e+08) | |
| Preventable Hospital Stays Rate | 4681.30 (1754.70) | 4530.76 (1769.74) | |
| Medicare Enrollment, Aged & Disabled Tot | 17732.60 (41185.96) | 16382.34 (39629.62) | 0.017 |
| Population | 96255.17 (2.5e+05) | 94521.84 (2.5e+05) | |
| Area | | | |
|   Non metro | 7,120 (63.3%) | 7,730 (66.7%) | <0.001 |
|   Metro | 4,130 (36.7%) | 3,858 (33.3%) | |
| Median Household Income | 48342.31 (10873.92) | 48822.27 (13634.26) | 0.003 |
| Total Number Hospitals | 2.00 (3.51) | 1.84 (3.14) | <0.001 |
| % Persons in Poverty | 15.30 (5.76) | 15.68 (6.30) | <0.001 |
| Unemployment Rate, 16+ | 6.00 (2.70) | 6.22 (2.99) | <0.001 |
| Boadband Tier 1 | | | |
|   Up to 200 | 122 (1.1%) | 201 (1.7%) | <0.001 |
|   201 to 400 | 1,719 (15.3%) | 2,727 (23.5%) | |
|   401 to 600 | 5,620 (50.0%) | 5,421 (46.8%) | |
|   601 to 800 | 3,515 (31.2%) | 2,764 (23.9%) | |
|   More than 800 | 274 (2.4%) | 475 (4.1%) | |
| Interstate Licensure Compact | 0.59 (0.49) | 0.28 (0.45) | <0.001 |

Total sample: N = 22,838

Mean (Standard deviation): p-value from linear regression.

Frequency (Percent%): p-value from Pearson test.

Main Results

Table A.C.I—: Effects of TPLs on Hospital Outpatient Visits and Log Medicare Costs

| | Hospital Outpatient Visits | | | Log Medicare Costs | | |
|---|---|---|---|---|---|---|
| | (1) | (2) | (3) | (4) | (5) | (6) |
| | Full sample | Non-metro | Metro | Full sample | Non-metro | Metro |
| **Post-Implementation Effects (Treated ×Post)** | | | | | | |
| Post Payment Parity | 0.011 | -0.017 | 0.019 | 0.057** | 0.058** | 0.041 |
| | (0.023) | (0.038) | (0.023) | (0.024) | (0.023) | (0.029) |
| Post Price Floor | 0.087** | 0.071** | 0.092** | 0.018 | 0.020 | 0.025* |
| | (0.035) | (0.034) | (0.041) | (0.011) | (0.014) | (0.014) |
| Post Price Parity | 0.032 | 0.104** | 0.014 | 0.031 | 0.026 | 0.050* |
| | (0.031) | (0.051) | (0.026) | (0.031) | (0.033) | (0.027) |
| Post Price Ceiling | -0.006 | -0.025 | -0.012 | -0.151*** | -0.171*** | -0.106*** |
| | (0.024) | (0.033) | (0.027) | (0.028) | (0.027) | (0.032) |
| Post Cost Parity | -0.134*** | 0.000 | -0.141*** | 0.045*** | 0.000 | 0.035* |
| | (0.019) | (.) | (0.025) | (0.016) | (.) | (0.018) |
| Post Cost Ceiling | -0.044** | 0.005 | -0.057** | -0.062** | -0.076*** | -0.027 |
| | (0.021) | (0.034) | (0.025) | (0.027) | (0.026) | (0.030) |
| **Controls** | | | | | | |
| Broadband | -0.005 | -0.400 | -0.006 | -0.076*** | -0.297** | -0.063*** |
| | (0.011) | (0.337) | (0.011) | (0.017) | (0.125) | (0.011) |
| Post Interstate Compact | 0.022 | 0.022 | 0.020 | 0.035* | 0.043** | 0.007 |
| | (0.020) | (0.039) | (0.020) | (0.018) | (0.018) | (0.019) |
| Log Population | 0.290 | 1.276** | 0.113 | 0.859*** | 0.789*** | 1.174*** |
| | (0.220) | (0.617) | (0.237) | (0.137) | (0.199) | (0.121) |
| Log Median HH Income | 0.059 | 0.037 | 0.079 | -0.038 | -0.031 | -0.086** |
| | (0.099) | (0.136) | (0.133) | (0.033) | (0.037) | (0.032) |
| Log Std Risk Adj Medicare Cost | -0.004 | 0.058 | -0.020 | 0.611*** | 0.604*** | 0.633*** |
| | (0.114) | (0.146) | (0.199) | (0.025) | (0.023) | (0.060) |
| Log Percent Poverty | -0.006 | -0.015 | 0.001 | 0.011 | 0.009 | 0.006 |
| | (0.032) | (0.053) | (0.038) | (0.018) | (0.021) | (0.020) |
| Log Unemployment Rate 16+ | -0.079* | -0.077 | -0.075 | 0.064** | 0.041 | 0.126*** |
| | (0.043) | (0.063) | (0.048) | (0.030) | (0.030) | (0.030) |
| Log % Age >= 65 w/o Health Insurance | 0.026 | -0.072* | 0.047* | 0.012 | 0.026 | -0.020 |
| | (0.022) | (0.042) | (0.024) | (0.020) | (0.025) | (0.018) |
| Time Fixed Effects | Yes | Yes | Yes | Yes | Yes | Yes |
| County Fixed Effects | Yes | Yes | Yes | Yes | Yes | Yes |
| Observations | 19077 | 12089 | 6988 | 22804 | 14816 | 7988 |

Standard errors in parentheses and clustered at state level
* $p < 0.10$, ** $p < 0.05$, *** $p < 0.01$

*Note I:* The table shows the difference-in-difference PPML estimates. The same specification has been followed in the subsequent tables. In subsequent tables, among controls only Broadband and Log Population are shown.

Table A.C.II—: Effects of TPLs on Medicaid Enrollment for Aged and Disabled and Preventable Hospital Stays Rate

| | Medicare Enrollment, Aged & Disabled | | | Preventable Hospital Stays Rate | | |
|---|---|---|---|---|---|---|
| | (1) | (2) | (3) | (4) | (5) | (6) |
| | Full sample | Non-metro | Metro | Full sample | Non-metro | Metro |
| **Post-Implementation Effects (Treated ×Post)** | | | | | | |
| Post Payment Parity | 0.009* | 0.022*** | 0.007 | 0.004 | 0.016 | -0.002 |
| | (0.005) | (0.008) | (0.005) | (0.014) | (0.011) | (0.056) |
| Post Price Floor | -0.004 | -0.014 | 0.004 | -0.052*** | -0.048*** | -0.036** |
| | (0.019) | (0.018) | (0.023) | (0.013) | (0.014) | (0.014) |
| Post Price Parity | 0.004 | 0.011 | 0.005 | 0.018 | 0.011 | 0.045 |
| | (0.006) | (0.011) | (0.005) | (0.032) | (0.031) | (0.041) |
| Post Price Ceiling | -0.013** | -0.040*** | -0.006 | -0.058*** | -0.055*** | -0.065 |
| | (0.005) | (0.008) | (0.006) | (0.015) | (0.011) | (0.054) |
| Post Cost Parity | 0.005*** | ___ | 0.004* | 0.101*** | ___ | 0.077*** |
| | (0.002) | | (0.002) | (0.014) | | (0.018) |
| Post Cost Ceiling | -0.007 | -0.026*** | -0.004 | 0.004 | -0.001 | -0.004 |
| | (0.005) | (0.009) | (0.005) | (0.017) | (0.014) | (0.050) |
| **Controls** | | | | | | |
| Broadband | -0.016*** | 0.006 | -0.015*** | 0.048*** | 0.435** | -0.002 |
| | (0.004) | (0.075) | (0.004) | (0.016) | (0.213) | (0.016) |
| Post Interstate Compact | 0.010*** | 0.007 | 0.011*** | -0.007 | -0.005 | 0.008 |
| | (0.004) | (0.007) | (0.004) | (0.015) | (0.016) | (0.016) |
| Log Population | 1.096*** | 0.743*** | 1.164*** | 0.089 | -0.085 | -0.240* |
| | (0.049) | (0.089) | (0.062) | (0.109) | (0.214) | (0.141) |
| Time Fixed Effects | Yes | Yes | Yes | Yes | | |
| County Fixed Effects | Yes | Yes | Yes | | | |
| Observations | 20502 | 13312 | 7190 | 18075 | 11691 | 6384 |

Standard errors in parentheses
* $p < 0.10$, ** $p < 0.05$, *** $p < 0.01$

*Note:* The table follows the same specification and uses the same controls as Table A.C.I. However, the rest of the controls are not shown.

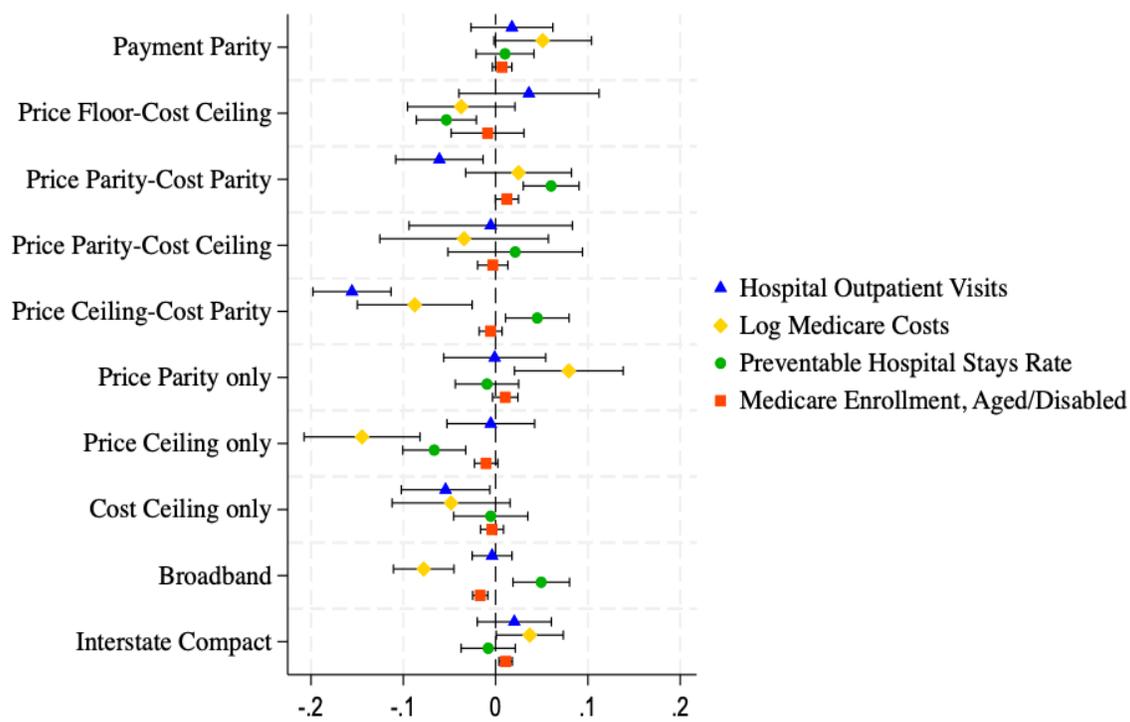

FIGURE A.C.I

Average Treatment Effect on the Treated (ATT) for Price Control-Cost Control types combinations

*Note*: The graphs shows the coefficient plots from table A.C.III. The X-axis represents actual coefficients. For treatment effect coefficients in percentage, the X-axis labels should be multiplied by 100.

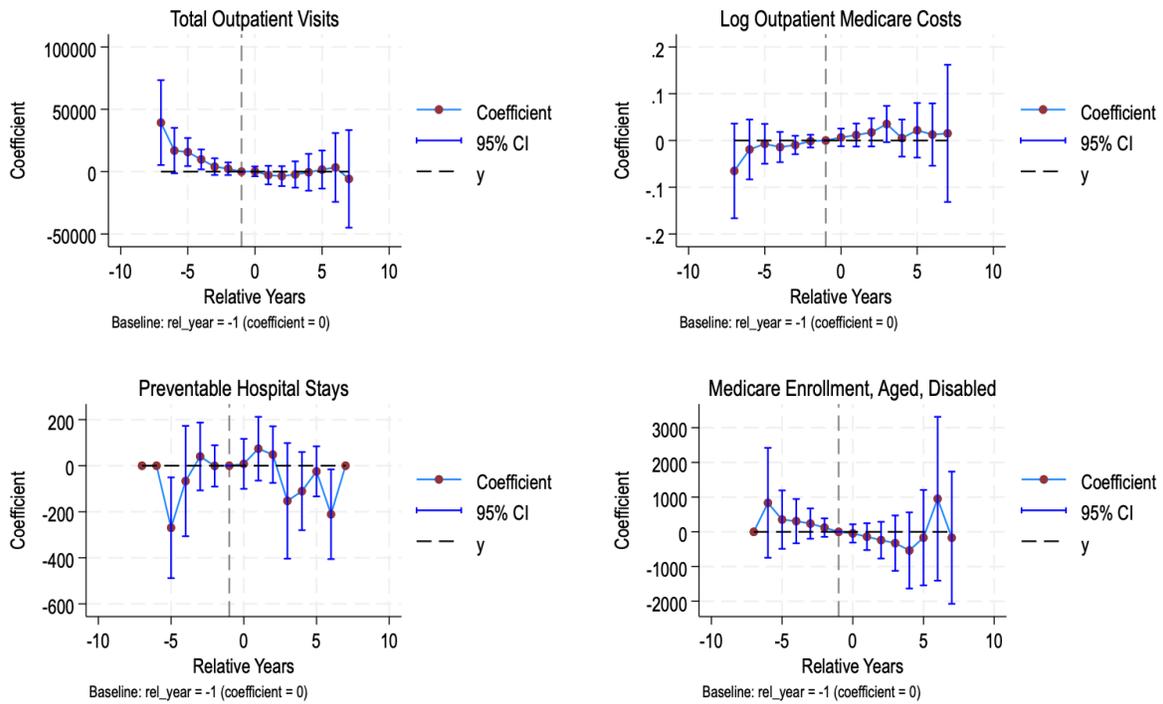

FIGURE A.C.II

Event Study Plots using Sun and Abraham (2021) Methodology

*Note*: We estimate average treatment effects for count outcomes (Outpatient Visits, Preventable Hospital Stays, Medicare Enrollment) using *ppmlhdfe* in our main specifications, which are designed for count data, and *reghdfe* for logged Outpatient Medicare costs. To explore dynamic treatment effects, we implement the Sun & Abraham method, a linear approach, across all outcomes. This unified framework accommodates staggered treatment adoption in our county-level data and simplifies visualization of effects over time. For count outcomes with large means, the linear SA method approximates dynamic effects reasonably well. To validate this, we conduct a separate event study using ppmlhdfe for Outpatient Visits, interacting treatment indicators with relative time periods. The consistency of dynamic patterns between this nonlinear model and the Sun & Abraham method confirms the latter's robustness, despite its linearity. Thus, the SA event study complements our main specifications by providing a practical and credible analysis of treatment dynamics.

Table A.C.III—: Effects of Various Factors on Hospital Outpatient Visits, Medicare Costs, Hospital Stays, and Medicare Enrollment, Aged & Disabled

| | (1) Hospital Outpatient Visits | (2) Log Medicare Costs | (3) Preventable Hospital Stays | (4) Medicare Enrollment Aged, Disabled |
|---|---|---|---|---|
| **Post-Implementation Effects (Treated ×Post)** | | | | |
| Post Payment Parity | 0.018 | 0.051* | 0.010 | 0.007 |
| | (0.023) | (0.026) | (0.016) | (0.005) |
| | | | | |
| Post Price Floor-Cost Ceiling | 0.036 | -0.037 | -0.053*** | -0.009 |
| | (0.039) | (0.029) | (0.017) | (0.020) |
| | | | | |
| Post Price Parity-Cost Parity | -0.061** | 0.025 | 0.060*** | 0.012* |
| | (0.024) | (0.028) | (0.015) | (0.006) |
| | | | | |
| Post Price Parity-Cost Ceiling | -0.005 | -0.034 | 0.021 | -0.003 |
| | (0.045) | (0.045) | (0.037) | (0.008) |
| | | | | |
| Post Price Ceiling-Cost Parity | -0.156*** | -0.088*** | 0.045** | -0.005 |
| | (0.022) | (0.031) | (0.018) | (0.006) |
| | | | | |
| Post Price Parity only | -0.001 | 0.079*** | -0.009 | 0.010 |
| | (0.028) | (0.029) | (0.017) | (0.007) |
| | | | | |
| Post Price Ceiling only | -0.005 | -0.145*** | -0.066*** | -0.010 |
| | (0.024) | (0.031) | (0.017) | (0.006) |
| | | | | |
| Post Cost Ceiling only | -0.054** | -0.048 | -0.005 | -0.004 |
| | (0.024) | (0.032) | (0.020) | (0.006) |
| | | | | |
| **Controls** | | | | |
| Broadband | -0.004 | -0.078*** | 0.049*** | -0.017*** |
| | (0.011) | (0.016) | (0.016) | (0.004) |
| | | | | |
| Post Interstate Compact | 0.020 | 0.037** | -0.008 | 0.011*** |
| | (0.020) | (0.018) | (0.015) | (0.004) |
| | | | | |
| Log Population | 0.295 | 0.856*** | 0.093 | 1.095*** |
| | (0.219) | (0.136) | (0.110) | (0.049) |
| Time Fixed Effects | Yes | Yes | Yes | Yes |
| County Fixed Effects | Yes | Yes | Yes | Yes |
| Observations | 19077 | 22780 | 18075 | 20502 |

Standard errors in parentheses
* $p < 0.10$, ** $p < 0.05$, *** $p < 0.01$

*Note:* The abbreviations used are: PF - Price Floor, PP - Price Parity, PC - Price Ceiling, CP - Cost Parity and CC - Cost Ceiling.

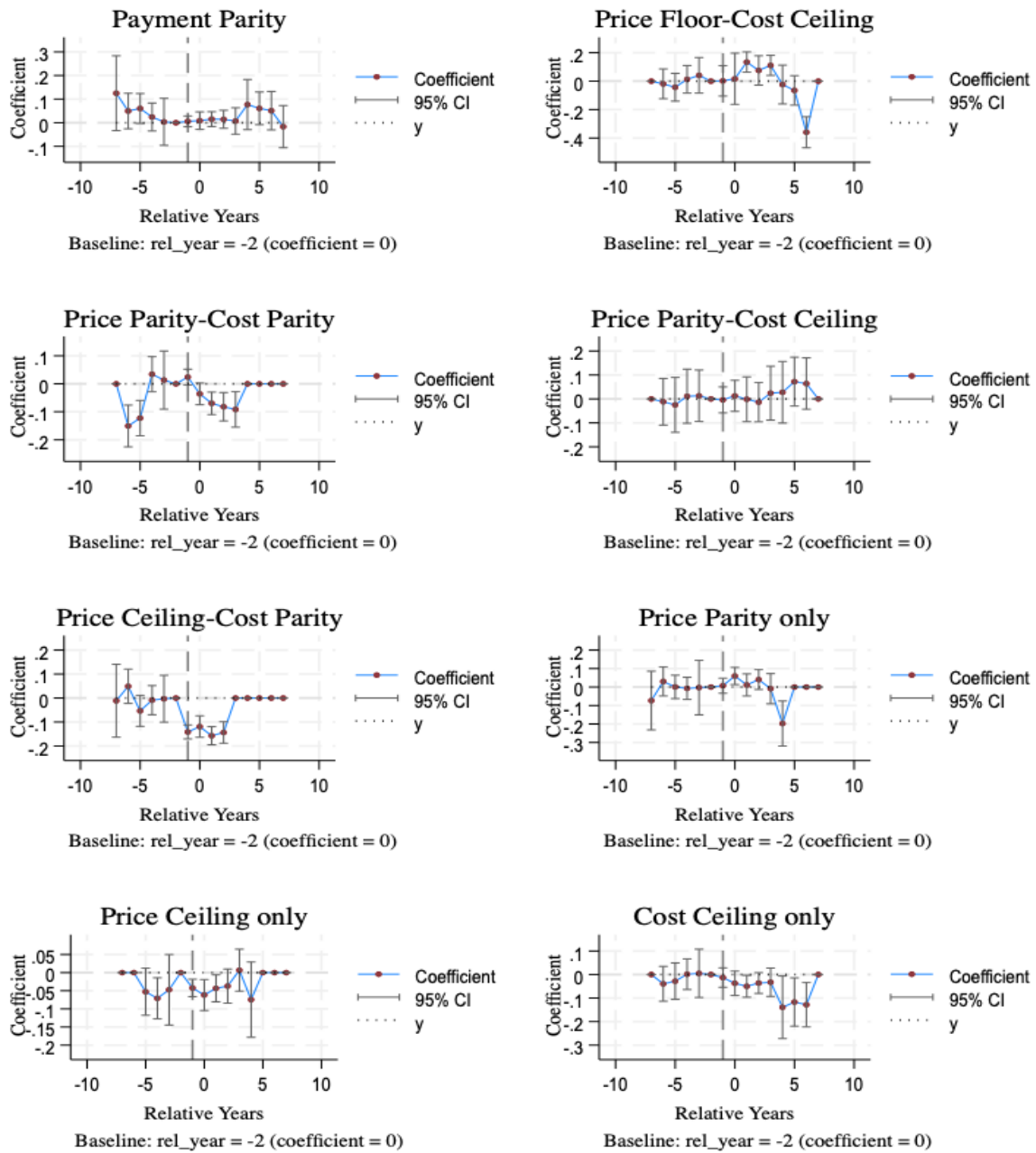

FIGURE A.C.III

Heterogeneous Treatment Effects by Policy Type for Total Hospital Outpatient Visits

*Note*: These plots show dynamic treatment effects of eight treatment types on Total Hospital Outpatient Visits, estimated using `ppmlhdfe` with a baseline at relative year = -2 (coefficient = 0). For the main treatment, Payment Parity (paypar), there is no anticipation at rel_year = -1 (95% CI includes 0), supporting the Conditional No Anticipation Assumption. Other treatment types, such as Price Ceiling-Cost Parity, exhibit minor anticipation at earlier pre-treatment years (e.g., rel_year = -3), but this does not impact the main analysis. The lack of anticipation in Payment Parity aligns with the ratio version of parallel trends from Chen and Roth (2024). For Price Ceiling-Cost Parity, the plot shows a sharp drop at rel_year = -1 and a sustained negative effect post-treatment (rel_year ≥ 0), consistent with the ATET of -15% from the main regression. Post-treatment coefficients (rel_year ≥ 0) represent dynamic treatment effects relative to rel_year = -2, while the overall policy impact (ATET) is the difference between the average post-treatment and pre-treatment differences.

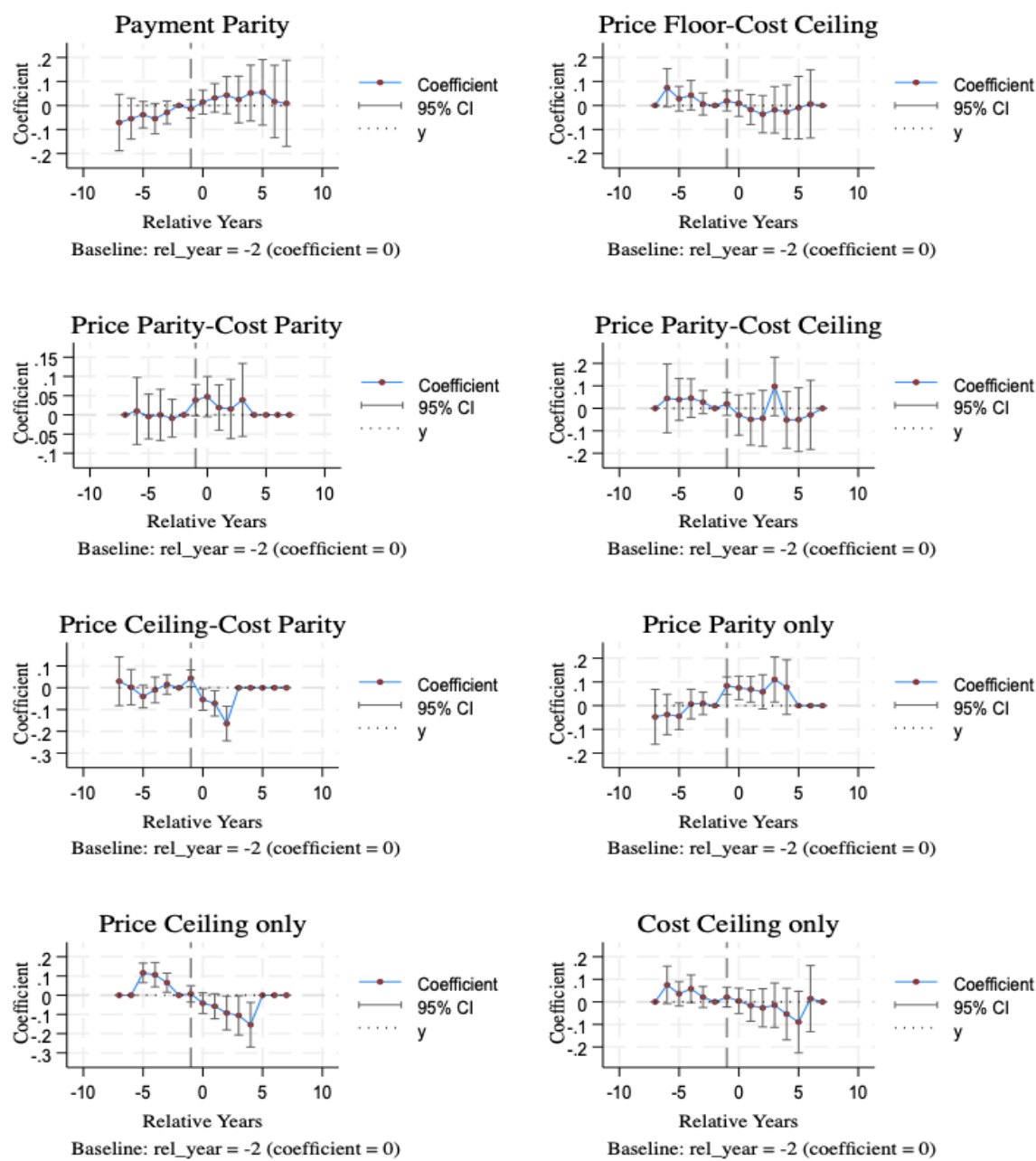

FIGURE A.C.IV

Heterogeneous Treatment Effects by Policy Type for Log Medicare Costs

*Note*: These plots show dynamic treatment effects of eight treatment types on Log Medicare Costs, estimated using `reghdfe` with a baseline at relative year = -2 (coefficient = 0). For the main treatment, Payment Parity (paypar), there is no anticipation at rel_year = -1 (95% CI includes 0), supporting the Conditional No Anticipation Assumption. Other treatments, such as Price Ceiling or Price Parity, exhibit minor anticipation at earlier pre-treatment years (e.g., rel_year = -3 for Price Ceiling), but this does not impact the main analysis. The lack of anticipation in Payment Parity aligns with the ratio version of parallel trends from Chen and Roth (2024). For Price Ceiling-Cost Parity, the plot shows a drop at rel_year = -1 and a sustained -8.8% effect post-treatment (rel_year $\geq$ 0), consistent with the ATET of -0.088 from the main regression. Post-treatment coefficients (rel_year $\geq$ 0) represent dynamic treatment effects relative to rel_year = -2, while the overall policy impact (ATET) is the difference between the average post-treatment and pre-treatment differences.

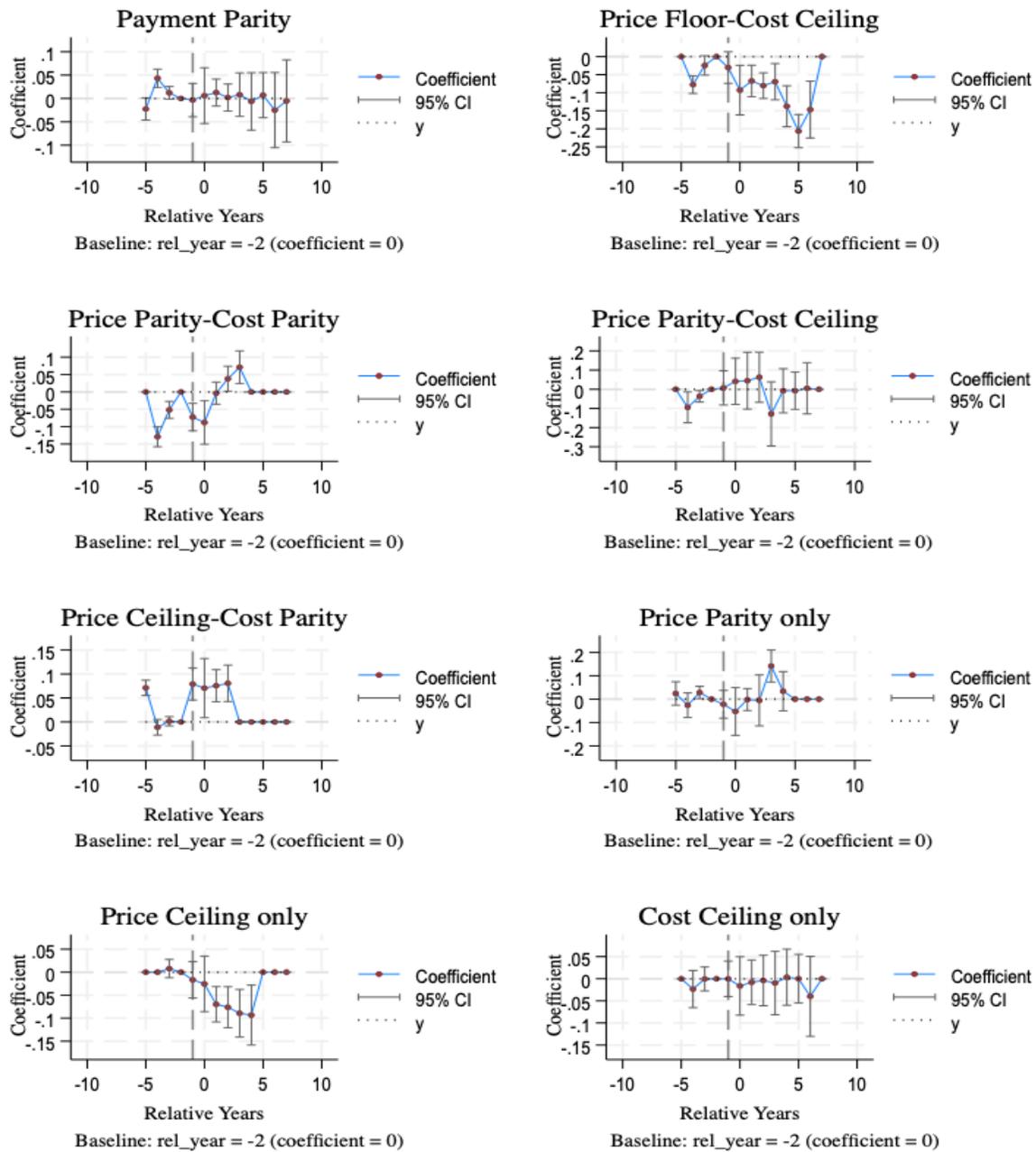

FIGURE A.C.V

Heterogeneous Treatment Effects by Policy Type for Preventable Hospital Stays

*Note*: These plots show dynamic treatment effects of eight policy types on Preventable Hospital Stays Rates (PHSR), estimated using `ppmlhdfe` with a baseline at relative year = -2 (coefficient = 0). For the main treatment, Payment Parity (paypar), there is no anticipation at rel_year = -1 (95% CI includes 0), supporting the Conditional No Anticipation Assumption. Other treatments, such as Price Parity-Cost Parity, exhibit minor pre-treatment fluctuations (e.g., rel_year = -4), but these are statistically insignificant (95% CI includes 0) and not sustained, thus not indicating anticipation. The lack of anticipation in Payment Parity aligns with the ratio version of parallel trends from Chen and Roth (2024). For Price Parity-Cost Parity, the plot shows a slight post-treatment increase, consistent with the ATET of 0.060 from the main regression. Post-treatment coefficients (rel_year ≥ 0) represent dynamic treatment effects relative to rel_year = -2, while the overall policy impact (ATET) is the difference between the average post-treatment and pre-treatment differences.

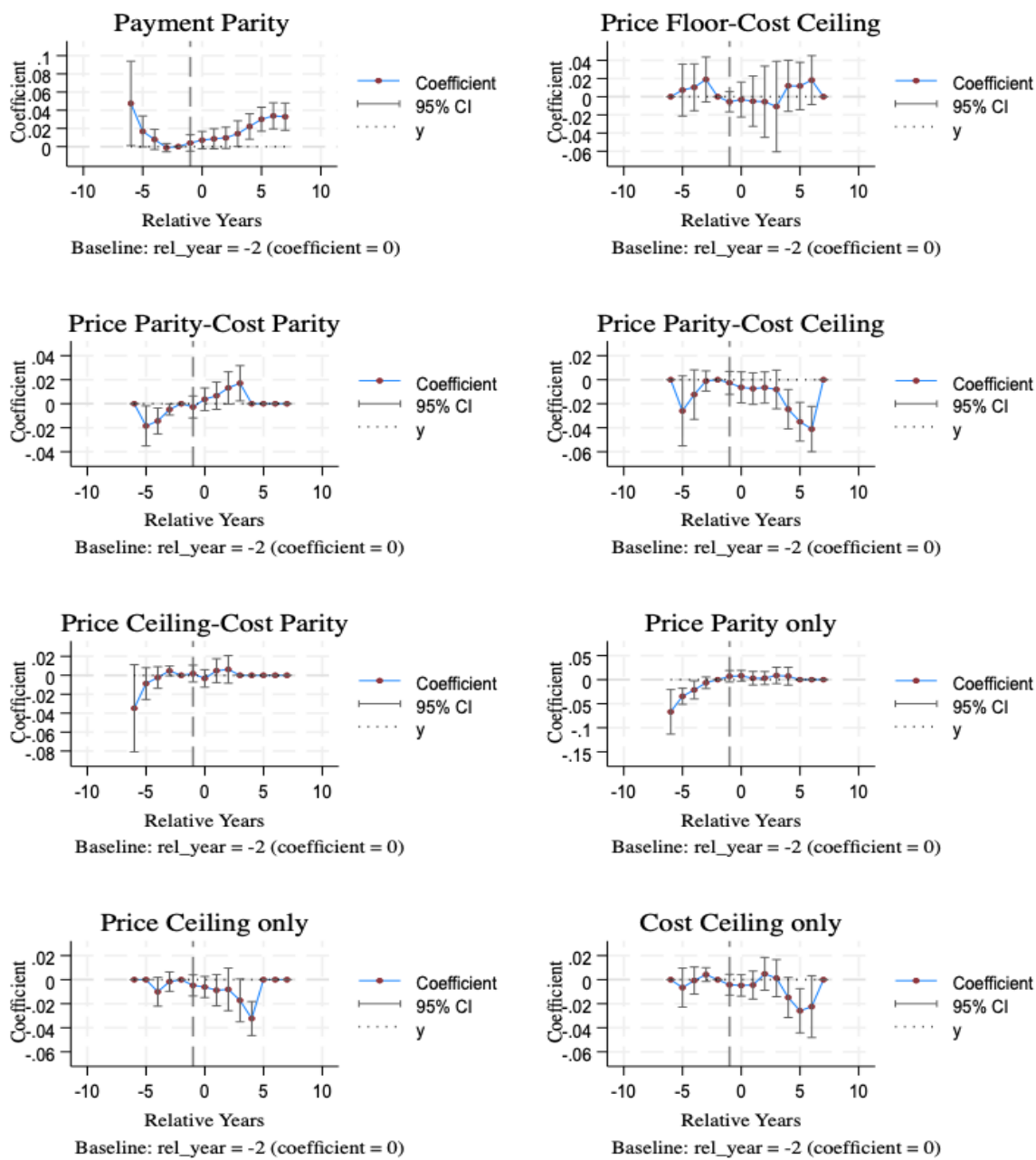

FIGURE A.C.VI

Heterogeneous Treatment Effects by Policy Type for Medicare Enrollment, Aged & Disabled

*Note*: This figure presents dynamic treatment effects of eight policy types on Medicare Enrollment for Aged and Disabled individuals, estimated using `ppmlhdfe` with a baseline at relative year = -2 (coefficient = 0). For Payment Parity (paypar), there is no anticipation at rel_year = -1 (95% CI includes 0), consistent with the Conditional No Anticipation Assumption. Price Ceiling-Cost Parity shows a significant drop at rel_year = -1 and a sustained negative effect post-treatment (rel_year $\geq$ 0), aligning with the ATET of -0.005 from the main regression. Minor pre-treatment fluctuations in other policies (e.g., rel_year = -3) are statistically insignificant (95% CI includes 0) and not sustained. The ATET reflects the overall policy impact, computed as the difference between average post-treatment and pre-treatment differences.

Table A.C.IV—: Estimates With Broadband Interaction

| | Without Considering Framing | | | Considering Framing | | |
|---|---|---|---|---|---|---|
| | (1) | (2) | (3) | (4) | (5) | (6) |
| | Full sample | Non-metro | Metro | Full sample | Non-metro | Metro |
| **Triple Interactions** | | | | | | |
| Post Payment Parity=1 × Broadband | -0.002 | 0.005 | -0.000 | 0.032* | 0.205 | 0.030 |
| | (0.003) | (0.256) | (0.003) | (0.018) | (0.393) | (0.021) |
| Post Price Floor=1 × Broadband | ___ | ___ | ___ | 0.008*** | 1.744*** | -0.007 |
| | | | | (0.003) | (0.624) | (0.004) |
| Post Price Parity=1 × Broadband | ___ | ___ | ___ | -0.036* | 0.813 | -0.030 |
| | | | | (0.020) | (0.661) | (0.022) |
| Post Price Ceiling=1 × Broadband | ___ | ___ | ___ | 0.020 | -0.151 | 0.023 |
| | | | | (0.027) | (0.417) | (0.032) |
| Post Cost Parity=1 × Broadband | ___ | ___ | ___ | -0.128*** | ___ | -0.130*** |
| | | | | (0.037) | | (0.038) |
| Post Cost Ceiling=1 × Broadband | ___ | ___ | ___ | -0.030 | -0.835* | -0.027 |
| | | | | (0.018) | (0.450) | (0.021) |
| **Post-Implementation Effects (Treated ×Post)** | | | | | | |
| Post Payment Parity | -0.008 | 0.017 | -0.019 | -0.022 | 0.019 | -0.026 |
| | (0.018) | (0.063) | (0.019) | (0.037) | (0.082) | (0.049) |
| Post Price Parity | | | | 0.069 | 0.220 | 0.052 |
| | | | | (0.044) | (0.158) | (0.051) |
| Post Price Floor | ___ | ___ | ___ | 0.080* | 0.412*** | 0.103** |
| | | | | (0.043) | (0.097) | (0.044) |
| Post Price Ceiling | | | | -0.015 | -0.060 | -0.017 |
| | | | | (0.044) | (0.085) | (0.057) |
| Post Cost Parity | | | | 0.067 | 0.000 | 0.072 |
| | | | | (0.061) | | (0.060) |
| Post Cost Ceiling | ___ | ___ | ___ | -0.010 | -0.139 | -0.015 |
| | | | | (0.040) | (0.086) | (0.054) |
| **Controls** | | | | | | |
| Broadband | 0.000 | -0.282 | -0.002 | 0.007 | 0.000 | 0.005 |
| | (0.012) | (0.344) | (0.012) | (0.019) | | (0.019) |
| Post Interstate Compact | 0.027 | 0.044 | 0.023 | 0.023 | 0.027 | 0.021 |
| | (0.020) | (0.042) | (0.021) | (0.022) | (0.042) | (0.022) |
| Log Population | 0.512** | 1.236** | 0.382* | 0.476** | 1.043* | 0.342 |
| | (0.201) | (0.567) | (0.205) | (0.218) | (0.607) | (0.217) |
| Time Fixed Effects | Yes | Yes | Yes | Yes | Yes | Yes |
| County Fixed Effects | Yes | Yes | Yes | Yes | Yes | Yes |
| Observations | 19077 | 12089 | 6988 | 19077 | 12089 | 6988 |

Standard errors in parentheses
* $p < 0.10$, ** $p < 0.05$, *** $p < 0.01$

*Note I:* The table shows the PPML estimates for the triple interaction. The dependent variable is the total hospital outpatient visits in a given county in a given year. Standard errors in parentheses are clustered at the state level. The first three columns show $ATT$ without accounting for the framing of TPLs, while the last three columns do account for the framing ($ATT_k$, where k is the treatment type). "___" indicates either no output or omitted output.

*Note II:* "Post Payment Parity" which represents the non-specific "Average Treatment Effect on the Treated" (ATT) is shown here, and has been included in all subsequent specifications.

Data Description

## I. *Area Health Resource file (AHRF) data description*

The sample of contains numerous health-related facets, including characteristics of the labor force such as the total count of individuals employed and unemployed who are aged 16 or older, and the rate of unemployment. Moreover, it provides poverty statistics, represented as the percentage of persons living in poverty, as well as important economic indicators like per capita personal income and median household income. Details of health insurance coverage segmented by different age groups and information pertaining to Medicare beneficiaries and costs are also included.

FIGURE A.D.I

Density Plots of Outcome Variables

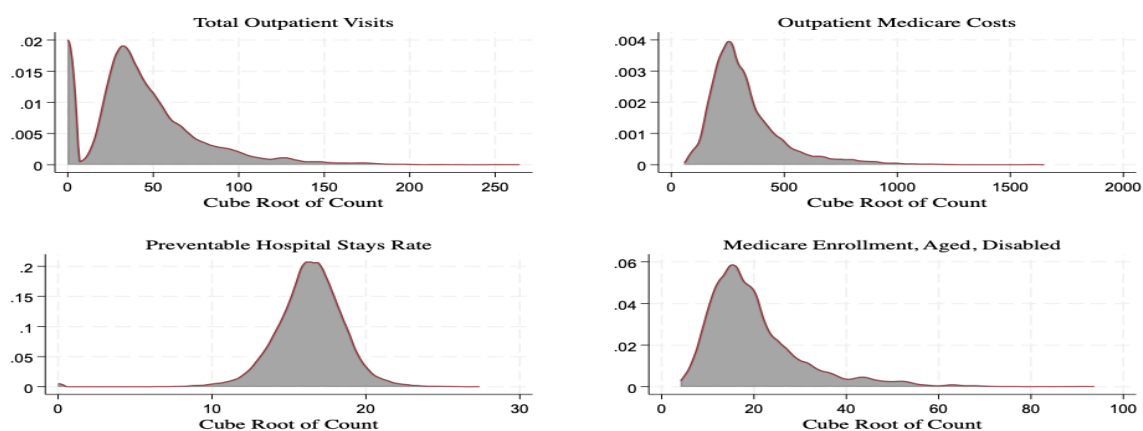

*Note*: The figure shows density plots for (in the clockwise direction): Total Hospital Outpatient Visits, Total Actual Medicare Costs, Medicare Enrollment (Aged and Disabled), Preventable Hospital Stays. Due to presence of count data for Total Hospital Outpatient Visits, Medicare Enrollment (Aged and Disabled), Preventable Hospital Stays and high prevalence of zeroes for Total Hospital Outpatient Visits, a direct visualization of the kernel density plots result in uninformative graphics with sharp spikes at zero and a horizontal line for values other than zero. The cube root transformation moderates the influence of outliers and spreads the data to allow for a subtler visualization of the distribution. Values on the x-axis represent the cube root of the number of physicians, with higher values indicating greater counts after transformation. The transformed values are plotted on the x-axis with the cube root of zero displayed at the origin. Non-zero values are spread out to allow for a more distinctive visualization of the distribution, enabling better identification of patterns and variations within the data.

The dataset supplies insights into hospital utilization rates across different ranges, data about inpatient days in various types of hospitals and nursing homes, and the total number of hospitals along with characteristics about each type of hospital. One of the most significant features of the data is the count of medical practitioners at a county level which is classified according to their type - whether they are Federal or Non-Federal, their field of specialty, their age group, their gender, and so on. The data also has 4 to 5 digit county FIPS codes consisting of 3 digit county code preceded by 1 or 2 digit State FIPS codes. To address missing values for "the percentage of people aged 65 and older without health insurance", for the years 2010–2012, a fixed-effects regression model was employed using Stata. The dataset was structured as panel data with countyfips as the panel identifier and year as the time variable. The variable was regressed on year for the period 2013 and onwards, controlling for unobserved heterogeneity and using robust standard errors. Predicted values were generated and used to impute the missing observations for 2010–2012, ensuring continuity and leveraging data from subsequent years for accurate interpolation.

FIGURE A.D.II

County-wise Degree of Urbanization and Bi-variate Distribution of Broadband Penetration and Physician Counts

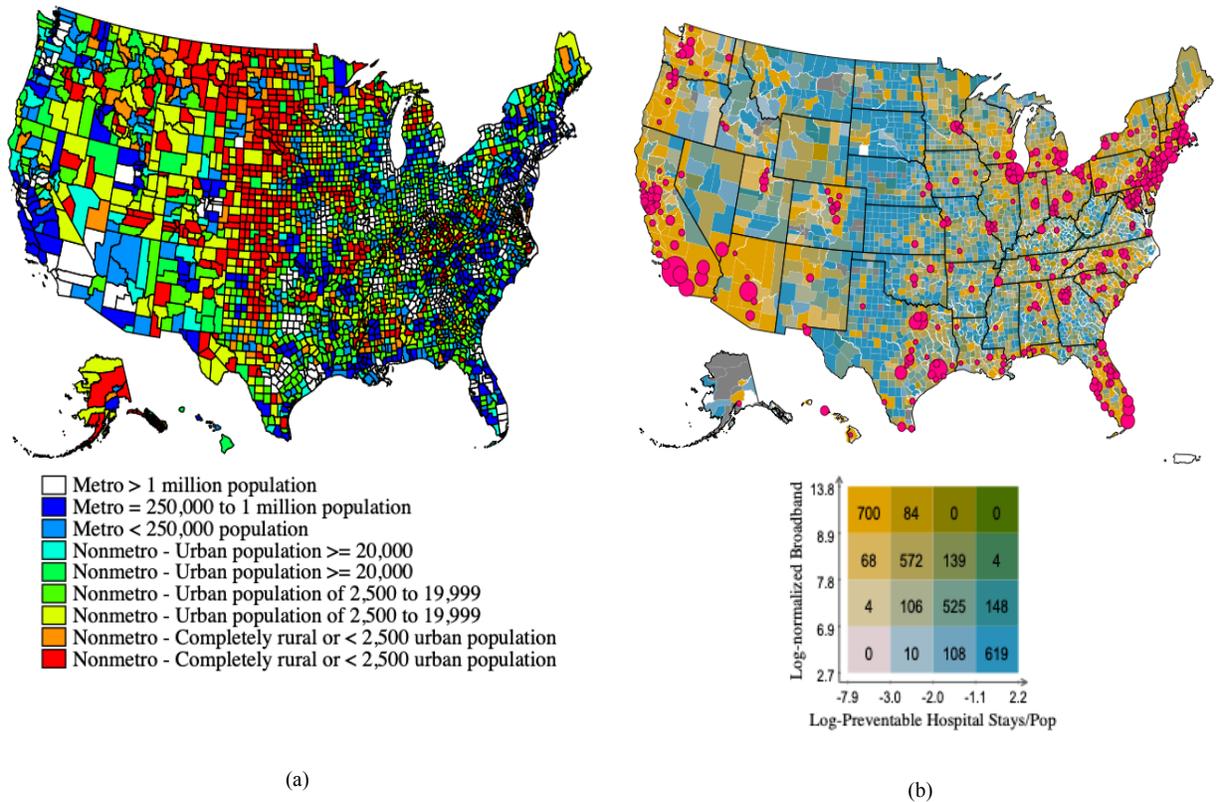

(a)

(b)

*Note*: Panel (a) shows the metro and non-metro, rural, and urban areas. The bluish shades correspond to the metro counties. The reddish-orange counties are non-metro rural and greenish-yellowish ones are non-metro urban. Panel (a) shows the bivariate distribution of broadband penetration and log of preventable hospital stays rate per population, for 2019. The Y-axis of the legend indicates log-normalized broadband score, while the X-axis of the legend indicates Total MDs. The numbers inside the legend box are the number of counties corresponding to each row-column intersection. The pink bubbles represent the densely populated areas with a population of more than 200,000, with the size of the bubbles proportional to the population of the county. The maps are generated using the Stata package from Naqvi (2023) using shapefiles from the U.S. Census Bureau (2016). In Panel(b), The metro areas (yellowish) mostly have high broadband penetration and low preventable hospital stays rate per population. The non-metro areas (bluish) mostly have low broadband penetration and high preventable hospital stays rate per population.

## II. Broadband Data Description

The dataset delineates four categorical variables, *Tier 1* through *Tier 4*, each accounting for the number of residential fixed broadband connections per 1000 households at different downstream speeds. Specifically, *Tier 1* covers connections with at least 200 Kbps, *Tier 2* includes those with a speed of 10 Mbps or more, *Tier 3* represents at least 25 Mbps, and *Tier 4* captures 100 Mbps and above. These Tiers are further subdivided into categories ranging from '0' to '5', each signifying a specific range of connections per 1000 households in each county for the respective speed tier. Category '0' denotes no connections, Category 1 signifies up to 200 household connections per 1000 households, Categories 2 to 4 represent 201-400, 401-600, and 601-800 household connections per 1000 households respectively, while Category 5 encompasses all situations where the connection exceeds 800 per 1000 households. This meticulous classification affords a detailed overview of U.S. household broadband connection distribution patterns. It is pertinent to note that Tier 1 inherently includes Tier 2.[24]

As defined by the FCC, broadband connections are lines (or wireless channels) that terminate at an end-user location and enable the end user to receive information from and/or send information to the Internet at information transfer rates exceeding 200 kilobits per second (kbps) in at least one direction.[25] Tier 1, which encompasses all other tiers and qualifies as "broadband" according to the FCC definition, provides a comprehensive measure of broadband residential connections and is available for both historical and current periods. Therefore, this study adopts Tier 1 as the measure of broadband penetration. The broadband tier data, sourced from the Federal Communications Commission, is expressed in terms of residential fixed broadband connections per 1,000 households.

To reconcile these measurements with the number of households represented in each dataset observation, a unique weighting variable, *hhweight*, was created. This variable was calculated by dividing the total number of households in each county by 1,000. The weighting factor was then applied to convert broadband connections into units compatible with the household counts. Specifically, the original *Tier 1* variable was multiplied by *hhweight*, yielding new weighted variables for each tier for accurate cross-county comparisons.

Subsequently, the weighted variables were standardized. The z-score transformation adds interpretabil-

---

[24]As per FCC, to assure firm confidentiality, the data regarding the 100 Mbps speed tier for the period from June 2014 to June 2016 has been redacted. The FCC also made a decision not to disclose data related to the 100 Mbps speed tier prior to December, 2016. The compilation of historical file (2008–2013) and the current file (2014 – present) employs different speed tiers. The sample is restricted to years 2009 to 2019. For instance, Tier 4 in the historical file includes connections with a downstream speed of at least 10 Mbps, while the same tier in the current file represents connections with a downstream speed of at least 100 Mbps. Therefore, to construct a consistent time-series analysis from 2009 to 2019 without data imputation, reliable usage of two speeds can be made: connections of at least 200 Kbps (Historical Tier 1 and Current Tier 1) and those of at least 10 Mbps (Historical Tier 4 and Current Tier 2). The four speed tiers are not mutually exclusive. All the connections accounted for in Tiers 2, 3, and 4 are inherently contained within Tier 1.

[25]For further details, please refer to https://transition.fcc.gov/form477/477glossary.pdf

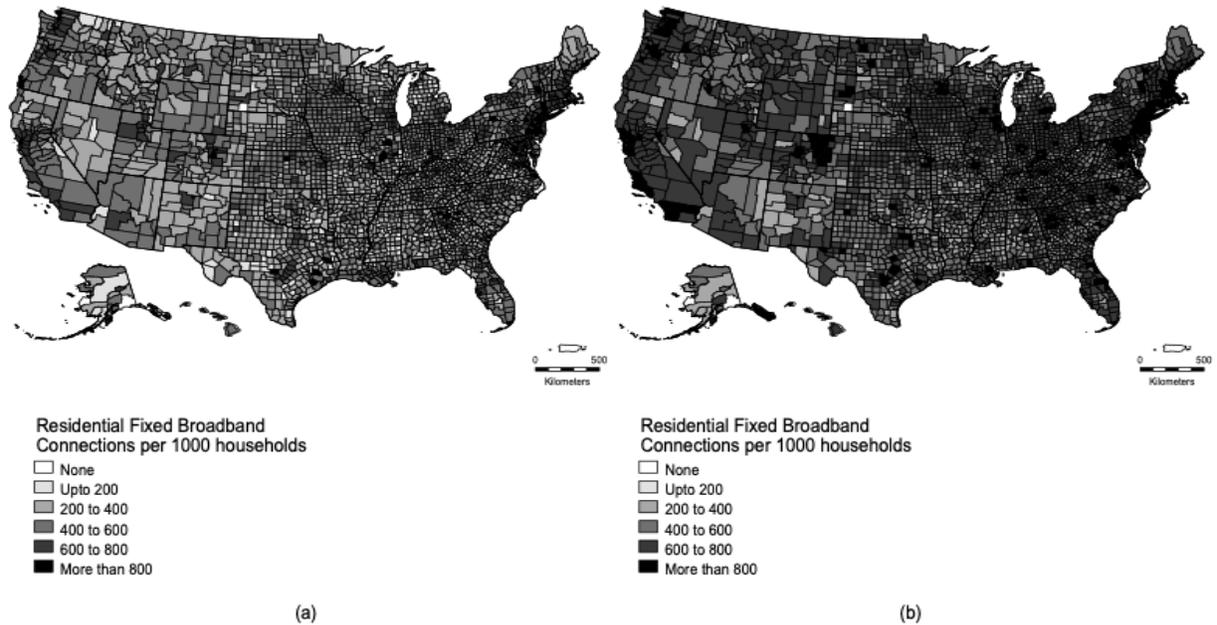

(a)

(b)

FIGURE A.D.III

Broadband Penetration

*Note*: Panel (a) and Panel (b) represent year 2010 and 2019 respectively. A visual comparison shows the increased broadband penetration owing to Federal and State level policy efforts. The Stata code used to generate all graphs and maps is available in the online supplementary file Figures_Script_SupplementaryData.do.

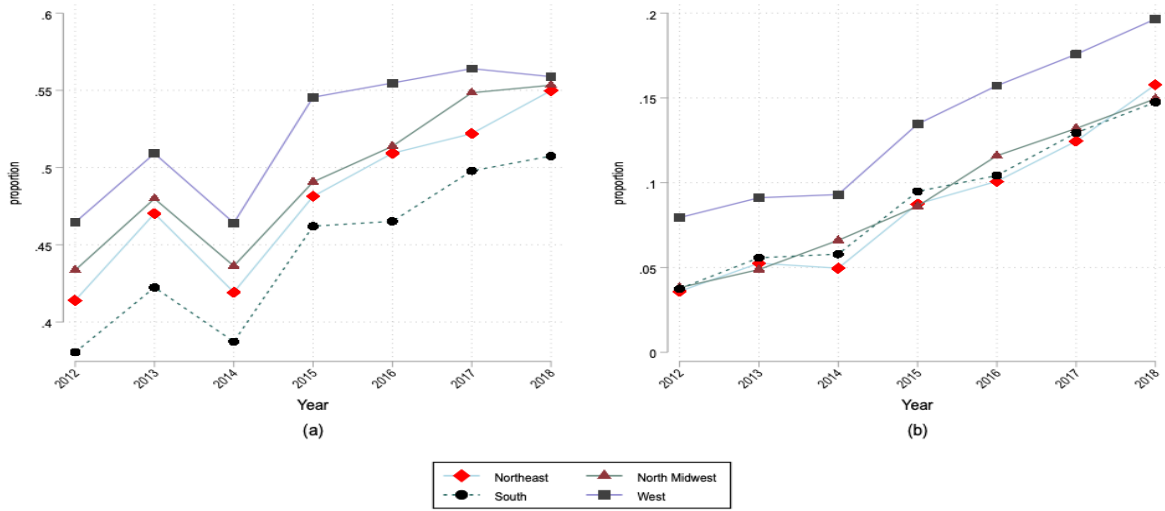

FIGURE A.D.IV

Health Related Usage of the Internet

*Note*: Panel (a) shows the region-wise proportion of people in the sample who "looked up health information on the internet in the past 12 months," and Panel (b) shows the region-wise proportion of people who "scheduled an appointment with a health care provider on the internet in the past 12 months." The data comes from the American Time Use Survey (Blewett et al., 2023).

ity by indicating whether an observation's value is above or below the mean and by how many standard units (standard deviations). Figure A.D.III illustrates the geographical differences in broadband penetration in 2010 and 2019. Although the county-level geographical disparity is evident in both panels, the increase in broadband penetration across the country from 2010 to 2019 is quite pronounced.

### III.   IMLC Data Description

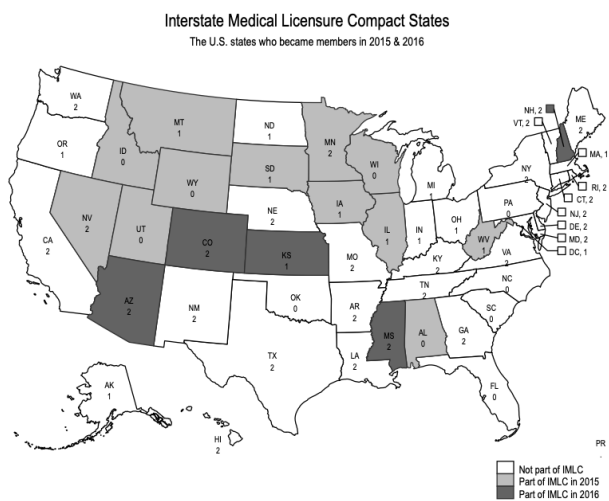

FIGURE A.D.V

Inter State Licensure Compact States

*Note*: The map indicates the states that joined the compact in 2015 and 2016. To accommodate the constraints of the map representation, Alaska and Hawaii are positioned below the contiguous United States rather than at their actual geographic locations. A more detailed exploration can be obtained from: https://www.imlcc.org/news/press-releases-and-publications/.

### Robustness

### Table A.E.I—: RESET Test Results

| | TotHospsOutpatientVisits | PreventableHospitalStaysRate | MedicareEnrollmentAgedDisa |
|---|---|---|---|
| Null Hypothesis ($H_0$) | | $(\hat{y})^2 = 0$ | |
| Test Statistic | | chi2(1) | |
| p-value | 0.3604 | 0.1118 | 0.1346 |

*Note*: The null hypothesis ($H_0$) in your context is that the model is correctly specified, meaning that the functional form of the model is appropriate. Specifically, it tests whether the model can be improved by adding higher-order terms of the predicted values (squared predicted values in this case). If the null hypothesis is not rejected, there is no evidence from the data to suggest that the model is misspecified. A low p-value indicates that we reject the null hypothesis, signifying model mis-specification. Here, for all models, the p-values are above the typical significance levels (0.1, 0.05, 0.01), and therefore the null hypothesis cannot be rejected. This suggests that there is no evidence of model misspecification.

### Table A.E.II—: Placebo Test Results for Outcome Variables

|  | Total Outpatient Visits | Preventable Hospital Stays | Medicare Enrollment, Aged, Disabled | Log Medicare Costs |
|---|---|---|---|---|
| *Treated × Relative years pre-fake-treatment* | -0.0043 | -0.0030 | 0.0016 | 0.0034 |
|  | (0.0040) | (0.0029) | (0.0008) | (0.0039) |
| Observations | 19,077 | 18,075 | 20,502 | 22,804 |

Standard errors in parentheses
* $p < 0.10$, ** $p < 0.05$, *** $p < 0.01$

The table presents the key results from the placebo test conducted to check for pre-treatment trends across four different outcome variables. A fictitious treatment period was assigned before the actual implementation of the treatment to observe any effects that should not exist if the parallel trends assumption holds. For all outcome variables, the coefficients for the interaction term (*Treated × Relative years pre-fake-treatment*) are not statistically significant. These results reinforce the validity of our main findings.

### Table A.E.III—: Pre-Trends Test Results for Outcome Variables

|  | Hospital Outpatient Visits | Preventable Hospital Stays | Medicare Enrollment, Aged, Disabled | Log Medicare Costs |
|---|---|---|---|---|
| *Treated × 7 years pre-treatment* | -0.1826 | ___ | ___ | -0.1729 |
|  | (0.4156) |  |  | (0.4431) |
| *Treated × 6 years pre-treatment* | -0.1847 | ___ | 0.0293 | -0.1322 |
|  | (0.3820) |  | (0.0431) | (0.4043) |
| *Treated × 5 years pre-treatment* | -0.1863 | 0.5230* | 0.0196 | -0.1216 |
|  | (0.3509) | (0.2893) | (0.0394) | (0.3648) |
| *Treated × 4 years pre-treatment* | -0.1847 | 0.4746* | 0.0119 | -0.1006 |
|  | (0.3179) | (0.2613) | (0.0361) | (0.3271) |
| *Treated × 3 years pre-treatment* | -0.1697 | 0.4165* | 0.0069 | -0.0817 |
|  | (0.2842) | (0.2320) | (0.0331) | (0.2846) |
| *Treated × 2 years pre-treatment* | -0.1608 | 0.3653* | 0.0048 | -0.0642 |
|  | (0.2506) | (0.2051) | (0.0286) | (0.2512) |
| *Treated × 1 years pre-treatment* | -0.1412 | 0.3266* | 0.0036 | -0.0461 |
|  | (0.2191) | (0.1756) | (0.0248) | (0.2167) |
| Observations | 9,407 | 9,092 | 10,377 | 11,554 |

Standard errors in parentheses
* $p < 0.10$, ** $p < 0.05$, *** $p < 0.01$
___ indicates omitted coefficients.

The table presents the key results from the pre-trends test conducted to check for parallel trends across four different outcome variables. The columns represent different outcome variables, and the rows represent the interaction of the treatment with relative years before the treatment. Coefficients that are statistically significant at the 10% level are marked with asterisks. The coefficients that are not statistically significant indicate that the parallel trends assumption cannot be rejected, supporting the validity of our main findings.

Table A.E.IV—: No Anticipation Test Results for Outcome Variables

| | Hospital Outpatient Visits | Preventable Hospital Stays | Medicare Enrollment, Aged, Disabled | Log Medicare Costs |
|---|---|---|---|---|
| *Treated × 7 years pre-treatment* | -0.0047 | -0.0089 | 0.0014 | -0.0008 |
| | (0.0034) | (0.0075) | (0.0021) | (0.0037) |
| *Treated × 6 years pre-treatment* | -0.0039 | -0.0065 | 0.0011 | -0.0010 |
| | (0.0029) | (0.0065) | (0.0018) | (0.0033) |
| *Treated × 5 years pre-treatment* | -0.0036 | -0.0061 | 0.0009 | -0.0012 |
| | (0.0025) | (0.0059) | (0.0016) | (0.0030) |
| *Treated × 4 years pre-treatment* | -0.0032 | -0.0057 | 0.0008 | -0.0011 |
| | (0.0023) | (0.0056) | (0.0015) | (0.0027) |
| *Treated × 3 years pre-treatment* | -0.0028 | -0.0052 | 0.0007 | -0.0009 |
| | (0.0022) | (0.0053) | (0.0013) | (0.0024) |
| *Treated × 2 years pre-treatment* | -0.0025 | -0.0049 | 0.0006 | -0.0008 |
| | (0.0020) | (0.0050) | (0.0012) | (0.0022) |
| *Treated × 1 year pre-treatment* | -0.0026 | -0.0050 | 0.0006 | -0.0007 |
| | (0.0018) | (0.0047) | (0.0011) | (0.0021) |
| Observations | 9,407 | 9,092 | 10,377 | 11,554 |

Standard errors in parentheses
* $p < 0.10$, ** $p < 0.05$, *** $p < 0.01$

The table presents the key results from the no anticipation test conducted to check for anticipatory effects across four different outcome variables. The columns represent lags of different outcome variables, and the rows represent the interaction between the relative pre-treatment years and the treatment indicator. Standard errors are clustered at the state level. For all outcome variables, the interaction terms are not statistically significant, indicating that the no anticipation assumption cannot be rejected. This means that the treatment effects observed post-treatment are likely not influenced by participants anticipating the treatment.